\documentclass[preprint,aps,prmaterials,groupedaddress,longbibliography]{revtex4-2} % change preprint to reprint or twocolumn to see two-col format

\usepackage{graphicx}
\usepackage{glossaries}
\usepackage{array,makecell}
\usepackage[export]{adjustbox} %for image align in table
\usepackage{booktabs}

\newcolumntype{L}[1]{>{\raggedright\let\newline\\\arraybackslash\hspace{0pt}}p{#1}}
\newcolumntype{C}[1]{>{\centering\let\newline\\\arraybackslash\hspace{0pt}}p{#1}}
\newcolumntype{R}[1]{>{\raggedleft\let\newline\\\arraybackslash\hspace{0pt}}p{#1}}
\begin{document}

\title{Metallization of colloidal crystals}

\author{Ali Ehlen}
\affiliation{Applied Physics Program, Northwestern University}
\author{Hector Lopez-Rios}
\affiliation{Department of Materials Science and Engineering, Northwestern University}
\author{Monica \surname{Olvera de la Cruz}}
\email{m-olvera@northwestern.edu}
\affiliation{Department of Materials Science and Engineering, Northwestern University}
\affiliation{Applied Physics Program, Northwestern University}

% declare acronyms
\newacronym{md}{MD}{molecular dynamics}
\newacronym{bcc}{BCC}{body-centered cubic}
\newacronym{fcc}{FCC}{face-centered cubic}
\newacronym{bct}{BCT}{body-centered tetragonal}
\newacronym{sh}{SH}{simple hexagonal}
\newacronym{sc}{SC}{simple cubic}
\newacronym{T}{$T$}{temperature}
\newacronym{nps}{NPs}{nanoparticles}
\newacronym{mit}{IMT}{insulator-metal transition}

\begin{abstract}

Colloidal crystals formed by size-asymmetric binary particles co-assemble into a wide variety of colloidal compounds with lattices akin to ionic crystals. Recently, a transition from a compound phase with a sublattice of small particles to a metal-like phase in which the small particles are delocalized  has been predicted computationally and observed experimentally. In this colloidal metallic phase, the small particles roam the crystal maintaining the integrity of the lattice of large particles, as electrons do in metals. A similar transition also occurs in superionic crystals, termed sublattice melting. Here, we use energetic principles and a generalized \gls{md} model of a binary system of functionalized \gls{nps} to analyze the transition to sublattice delocalization in different co-assembled crystal phases as a function of \gls{T}, number of grafted chains on the small particles, and number ratio between the small and large particles $n_s$:$n_l$. We find that $n_s$:$n_l$ is the primary determinant of crystal type due to energetic interactions and interstitial site filling, while the number of grafted chains per small particle determines the stability of these crystals. We observe first-order sublattice delocalization transitions as \gls{T} increases, in which the host lattice transforms from low- to high-symmetry crystal structures, including A20 $\rightarrow$ BCT $\rightarrow$ BCC, A$_\mathrm{d}$ $\rightarrow$ BCT $\rightarrow$ BCC, and BCC $\rightarrow$ BCC/FCC $\rightarrow$ FCC transitions and lattices. Analogous sublattice transitions driven primarily by lattice vibrations have been seen in some atomic materials exhibiting an insulator-metal transition also referred to as metallization. We also find minima in the lattice vibrations and diffusion coefficient of small particles as a function of $n_s$:$n_l$, indicating enhanced stability of certain crystal structures for $n_s$:$n_l$ values that form compounds.

\end{abstract}


\maketitle

\section{Introduction}

Binary colloids of size-asymmetric particles have been shown to co-assemble into a diverse set of binary crystals \cite{Leunissen2005,Shevchenko2006,Yi2013,Eldridge1993,Cherniukh2021,Dinsmore1998,Filion2009a,Travesset2017a}. These crystals are compounds akin to atomic ionic crystals because the smaller particles occupy interstitial sites of a lattice formed by the large particles. Recently the exploration of binary colloidal crystals with highly size-asymmetric functionalized \gls{nps} has yielded the observation of crystal assemblies where the small \gls{nps} delocalize, rather than remaining fixed at interstitial sublattice sites \cite{Girard2019,Lopez-Rios2021,Cheng2021}. This phenomenon was also observed in simulations of colloidal crystals of oppositely charged, highly size-asymmetric, and highly charge-asymmetric nanoparticles with screened Coulomb interactions \cite{Lin2020,Tauber2016}.  In all these systems, the delocalized and diffusive small particles keep the large particles in fixed lattice positions, as electrons do in crystalline metals. The result is a metal-like colloidal crystal. 
 
The degree of sublattice delocalization was quantified using a normalized Shannon entropy, termed metallicity, by Girard and Olvera de la Cruz \cite{Girard2018,Girard2019}. They used simulations of co-assembled DNA-functionalized \gls{nps} that were highly asymmetric in size and grafting density of complementary linkers. These showed that sublattice delocalization, and consequently metallicity, increased with \gls{T}, changing the crystal from ionic to metallic. Furthermore, Girard and Olvera de la Cruz discovered a minimum in metallicity as a function of the ratio of the number of small \gls{nps} ($n_s$) to the number of large \gls{nps} ($n_l$) in the crystal. They used simple band structure construction concepts from solid state physics to explain the observed minimum in metallicity and equated metallicity to conductivity in metals \cite{Girard2018}. In this analogy, the value $n_s$/$n_l$ is the ``valency," and the metallicity, akin to conductivity, decreases with increasing $n_s$/$n_l$ as interstitial sites are filled until it reaches a minimum at the compound values of the lattice, when the interstitial sites are saturated (\textit{i.e.}, $n_s$/$n_l = 6$ for a \gls{bcc} crystal). Upon further increase of $n_s$/$n_l$, the metallicity increases as the conductivity does in atomic systems with increasing number of electrons in the conduction band. They also highlighted that the minimum in metallicity becomes sharper with an increase in the interaction energy between the small and large \gls{nps}, achieved by increasing the number of linkers on the small \gls{nps}. They also suggested that the localization-delocalization transition in colloidal crystals can be described as a classical analog to a Mott-like \gls{mit} in atomic systems. 

Interestingly, sublattice delocalization is also observed in non-metallic atomic systems, specifically superionic materials \cite{Hull2004}, and the transition to superionic sublattice delocalization is often termed ``sublattice melting." A canonical superionic material is AgI, in which the larger atomic species I forms a \gls{bcc} host lattice through which Ag ions diffuse. The Ag ions have been identified as diffusing between neighboring \gls{bcc} tetrahedral sites \cite{Schommers1977,Schommers1980}, and diffusion has been seen to be strongly coupled to the dynamics of the host lattice \cite{Salamon1979,Brenner2020}. 

Recently, we have observed similar behavior in colloidal systems by using a generalized \gls{md} model of a binary, size-asymmetric system of functionalized \gls{nps} with $n_s$:$n_l =$ 6:1. We reported the formation of stable colloidal \gls{bcc} crystals with a diffusive sublattice of small particles translating between neighboring tetrahedral sites \cite{Lopez-Rios2021}. Similar to AgI, we observed a strong correlation between diffusion and lattice vibrations as a function of \gls{T}, but we noted that the transition to sublattice delocalization is described by a smooth change, rather than a true phase transition. This suggests that phonons play an important role in the delocalization transition, and that an atomic analog to this classical localization-delocalization transition should include the effect of the interactions of the phonons with metallic electrons as in the Peierls \gls{mit}.  

\begin{figure}
    \includegraphics[width=3.75in]{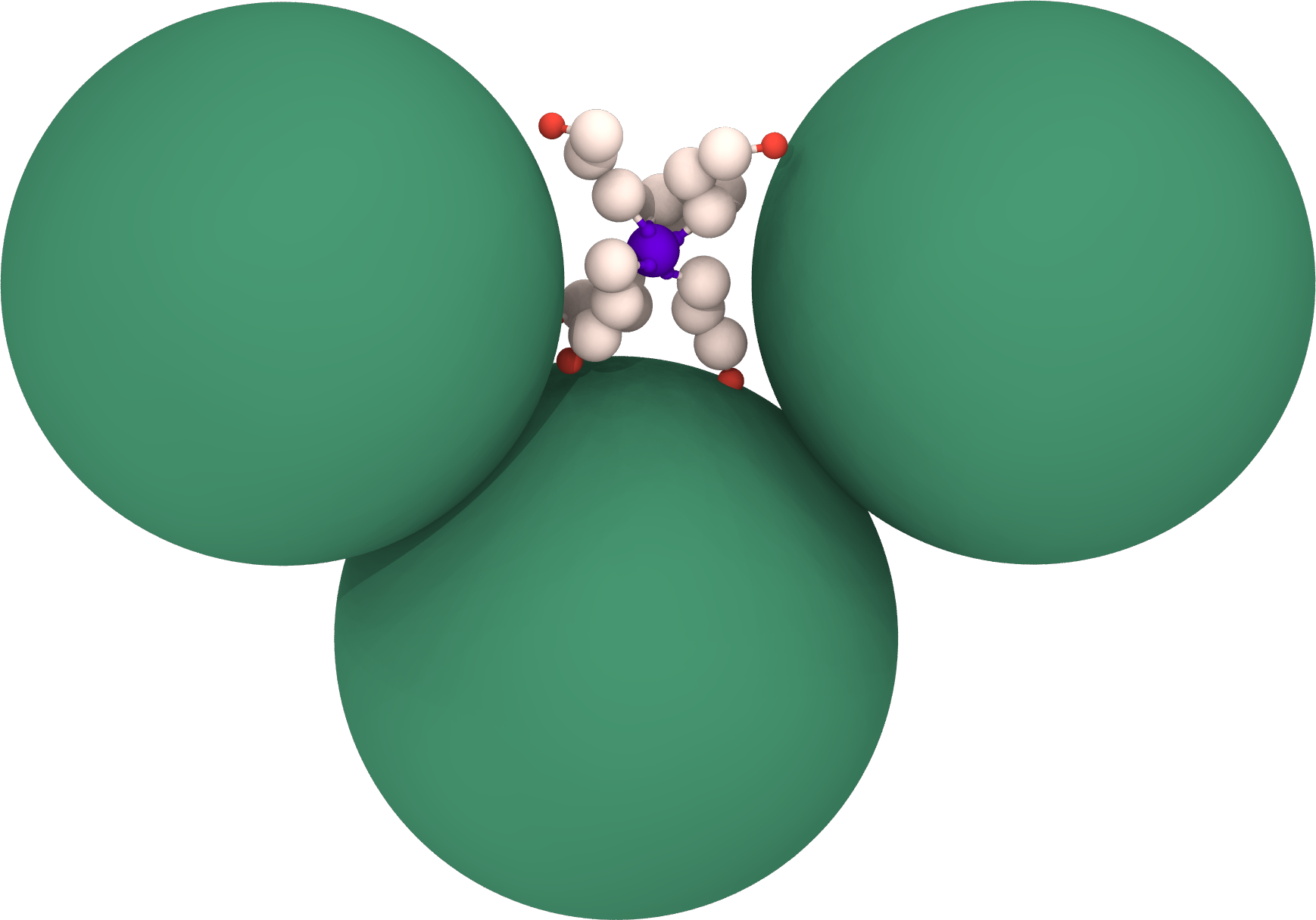}
    \caption{\label{fig:model} Graphical representation of the \gls{md} model. All beads have excluded volume interactions with each other, and there is an attractive interaction between the large particles (turquoise) and the interactive end of each chain (orange), as described in \cite{Lopez-Rios2021}. Note that there is no attractive interaction between large particles. Therefore, assemblies of these particles are held together exclusively by the attraction between the small particles' interactive beads and the large particles. Because the large particles represent densely grafted large particles, some overlap is permitted.}
\end{figure}

Here, we study the transition to sublattice delocalization at different values of the number ratio $n_s$:$n_l$, as a function of \gls{T} and the number of grafted chains per small particle, and we examine the origin of the delocalization transition. We highlight the similarities with the \gls{mit} and with superionic sublattice melting and analyze the effect of the phonons in the localization-delocalization transition. We use the \gls{md} model established in \cite{Lopez-Rios2021} in the NPT ensemble at near zero pressure to ensure that the resulting assemblies are due to interactions between small and large particles alone. The model, consisting of mutually attractive and size-asymmetric \gls{nps}, is visually depicted in Fig. \ref{fig:model}. The turquoise sphere is a coarse-grained representation of a large particle with either densely grafted chains or a functionalized surface. The small particle is represented by a central sphere (purple) and explicitly modeled grafted chains (white), each of which has an interactive terminus (orange) that is radially attractive only to large particles. The generality of the model implies that we can represent a variety of experimental systems \cite{Cherniukh2021,Dinsmore1998,Demirors2013,Sacanna2010,Lee2020,Lu2015}, and the tunability of \gls{nps} enables us to find a rich variety of lattices and multiple types of delocalization transitions. 
 
Using this model, we find that the crystal structure is determined by $n_s$:$n_l$ and the lattice stability is determined by the number of grafted chains per small particle. We observe a variety of crystals, including A20 and \gls{bct} lattices, and we confirm that the low \gls{T} (localized sublattice) positions of the small particles can be understood by analyzing their potential energy landscape. Almost all studied systems undergo a transition to sublattice delocalization with increasing \gls{T}, and the type of transition is also determined by $n_s$:$n_l$ based on energetic interactions and interstitial site filling. For some $n_s$:$n_l$ ratios, the sublattice smoothly delocalizes without undergoing a phase transition. This occurs for cubic lattices with nearly or completely full sublattice sites, near 6:1 and 10:1. For other number ratios, we observe a first-order sublattice delocalization transition accompanied by a first-order host lattice transition to a crystal of higher symmetry with inherent sublattice vacancies. This is seen in transitions from A20 to \gls{bct}, \gls{bct} to \gls{bcc}, and \gls{bcc} to \gls{fcc}, which all occur upon increasing \gls{T}. We present evidence that these transitions are entropic and driven by lattice vibrations, similar to the metallization of atomic materials driven by phonons, as in the Peierls \gls{mit} \cite{Budai2014}. Finally, we identify minima in the lattice vibrations and diffusion coefficient of the small particles as a function of $n_s$:$n_l$. Crystals at the minima are those whose interstitial sites are saturated with small particles, except the high-$n_s$:$n_l$ \gls{fcc} crystals. 
 
This article is organized as follows. In the next section, we will describe the range of crystal lattices observed in our parameter space of 4, 6, 8, and 10 grafted chains per small particle and number ratios $n_s$:$n_l$ between 3:1 and 10:1, over a wide range of temperatures. We will then further detail the three delocalization behaviors we observe and discuss the implications of the diffusion coefficient minima.

\section{Results}

\subsection{Determining crystal structure by number ratio $n_s$:$n_l$}
\label{sect:compositions}

At low temperatures, the large particles form a variety of lattices with the small particles localized at interstitial sites. These sites are always Wyckoff positions, which have a unique set of symmetry operators associated with the host lattice. The location of the small particles at these interstitial sites is dependent only on crystal type. We find that the symmetry of the resulting lattices depends on $n_s$:$n_l$, and the stability of the lattice depends on the number of chains per small particle. 

\begin{table} 
    \begin{tabular}[t]{L{0.75in} R{1.65in} L{0.05in} L{0.85in} L{1.2in} L{1.55in}}
     \hline
     \multicolumn{2}{L{2.4in}}{Lattice type, space group } & & \# lattice pts/unit cell &
       Wyckoff position \newline (\# NNs/site) &
       $n_s$:$n_l$ ratios that result in this lattice \\
    \hline 
      A20 \newline 63 Cmcm  &
          {\fontsize{10}{12}\selectfont (3:1)}
          \includegraphics[width=0.75in,valign=t]{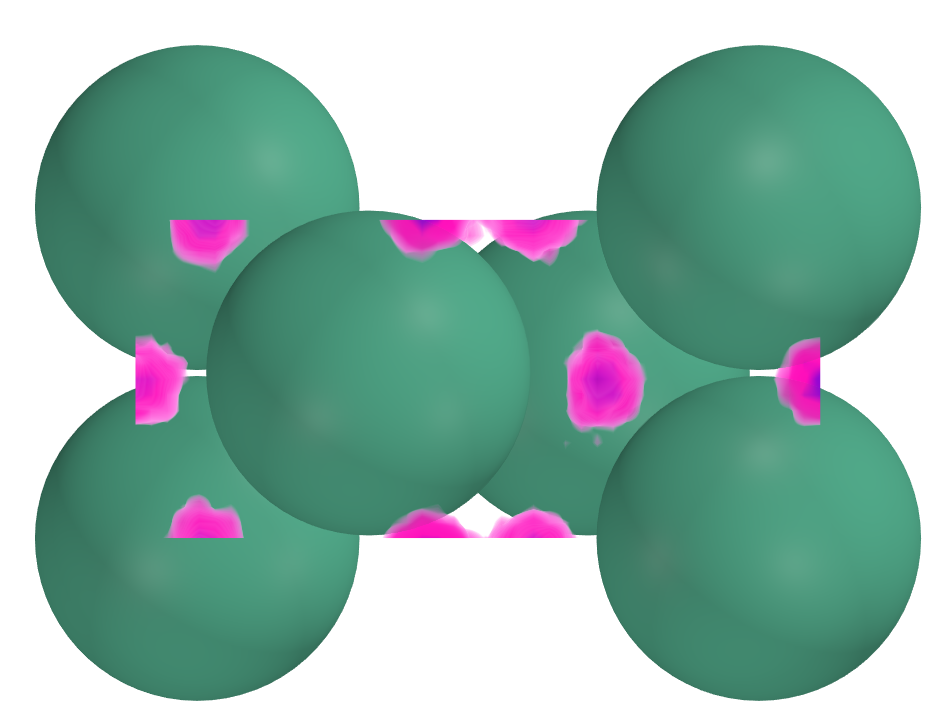} 
          \includegraphics[width=0.5in,valign=t]{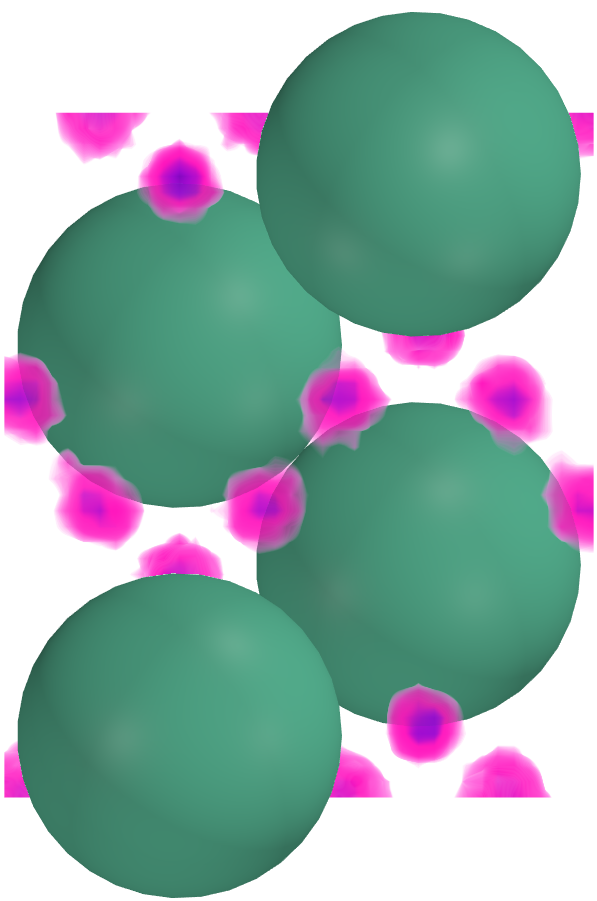} \newline
          {\fontsize{10}{12}\selectfont (4:1)}
          \includegraphics[width=0.75in,valign=t]{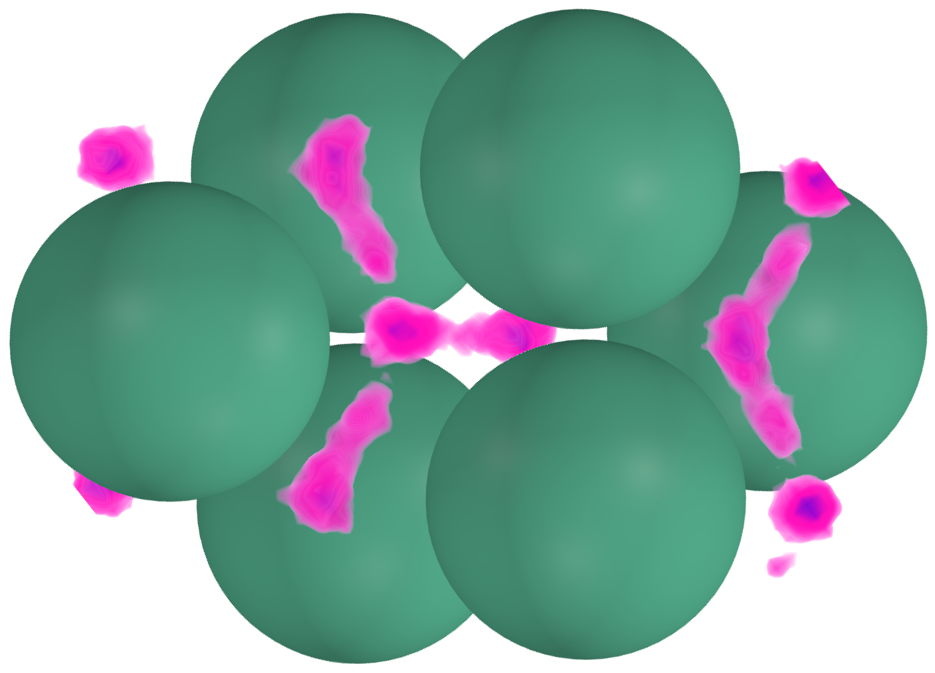} 
          \includegraphics[width=0.5in,valign=t]{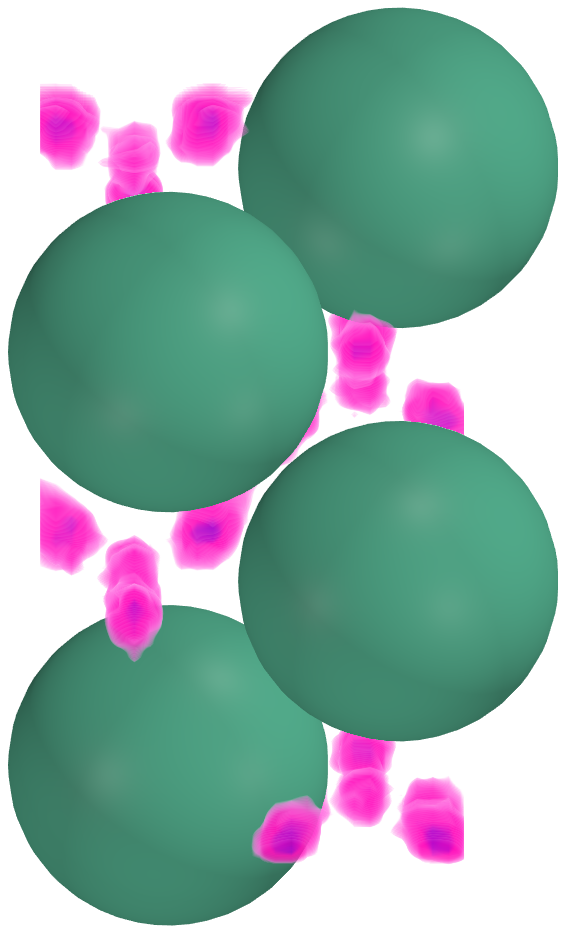} 
      & & 4 & 
      16h, 4c (3:1 systems) or \newline 
      16h, 8g ($\times 2$) \newline (4:1 systems) \newline (all 5 NNs)&
      3:1 and 4:1, resulting in different parameter ratios  \\
    \hline
      $\mathrm{A}_{\mathrm{d}}$ \newline  129 P4/nmm  &
          \includegraphics[width=0.6in,valign=t]{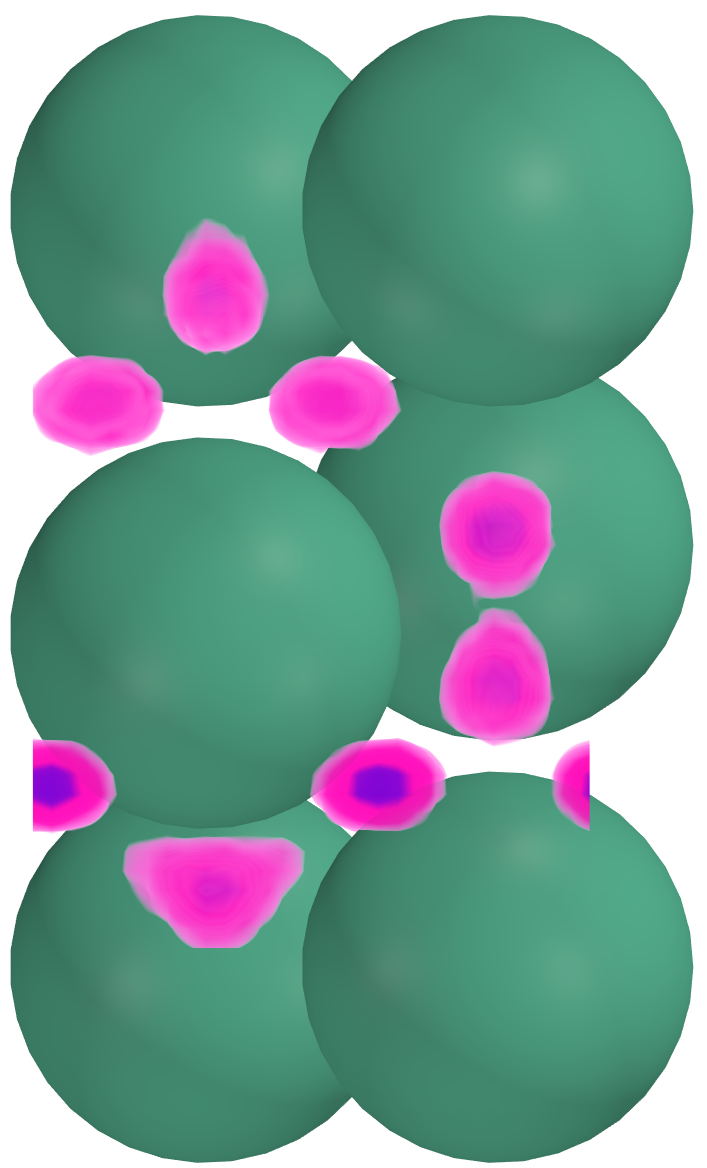} 
          \includegraphics[width=0.6in,valign=t]{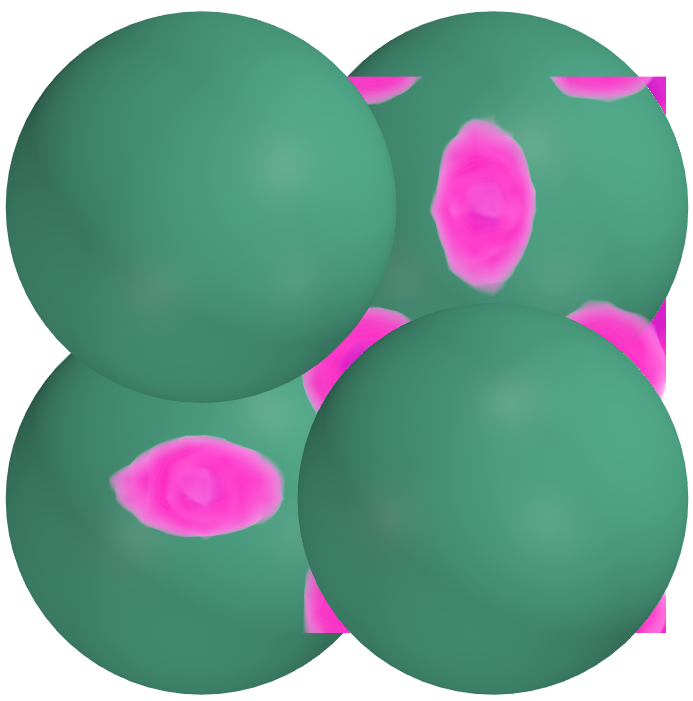} 
      & & 4 & 
     2c ($\times 2$), 4f, 8j (4-5 NNs) &
      4:1  \\
    \hline
      BCT \newline  139 I4/mmm &
      \includegraphics[width=0.75in,valign=t]{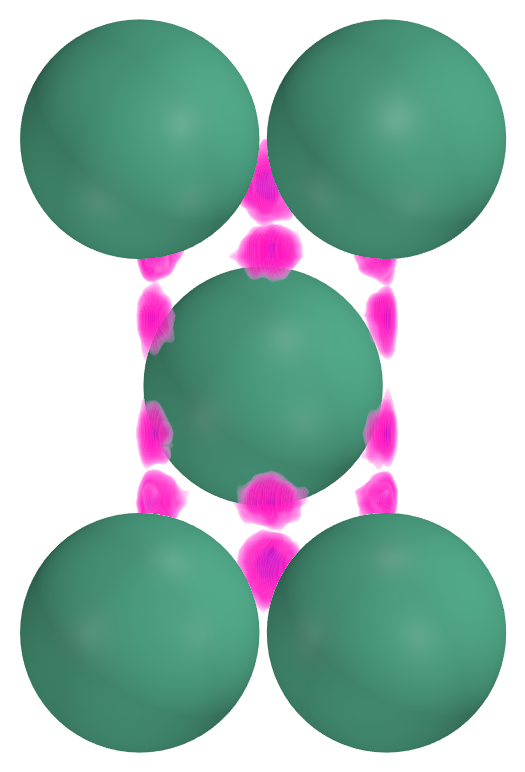} 
      \includegraphics[width=0.75in,valign=t]{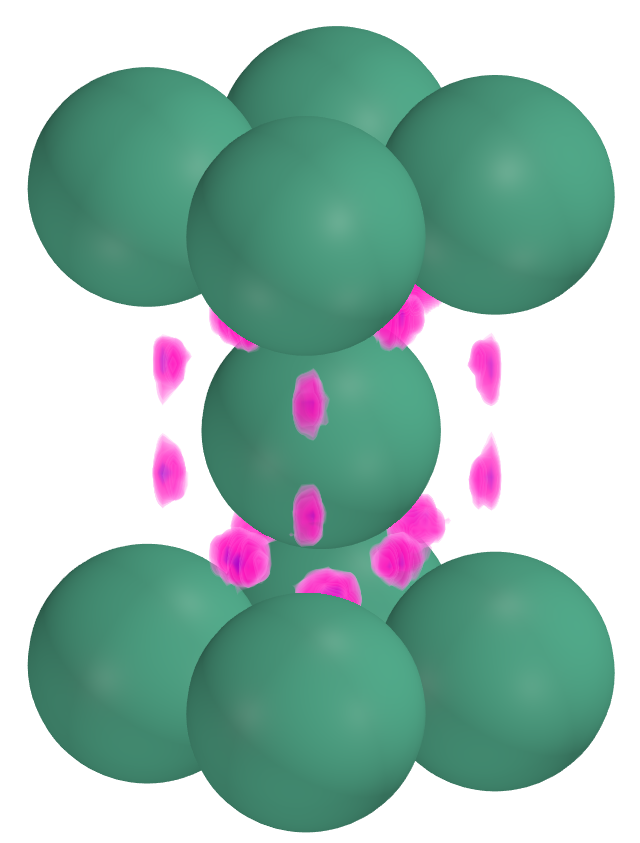} 
      & & 2 &
      4d (4 NNs)\newline 4e (5 NNs) &
      4:1 (with $c/a =2$, as shown here)\\
    \hline
      BCC \newline 229 Im$\overline{3}$m &
      \includegraphics[width=0.75in,valign=t]{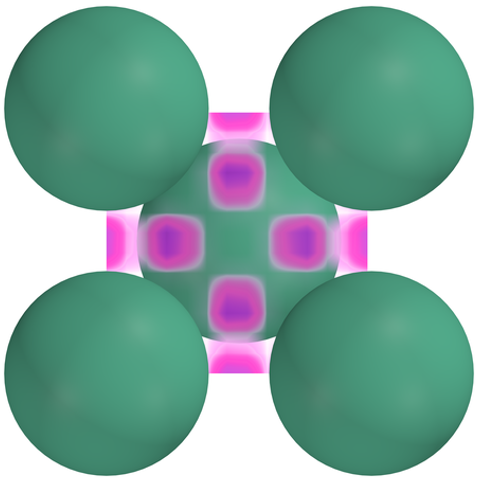} 
      \includegraphics[width=0.75in,valign=t]{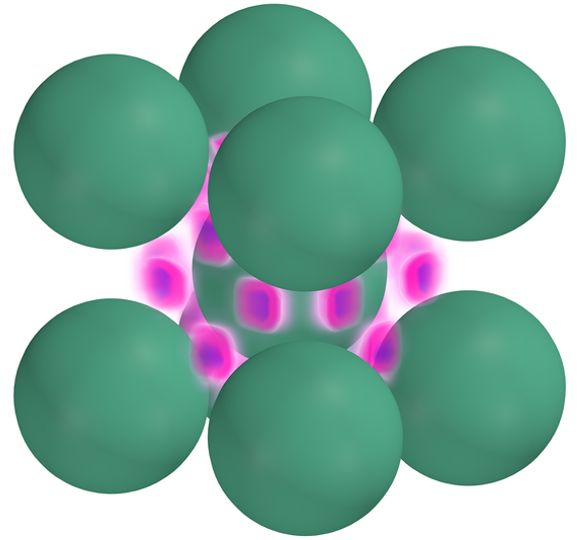}
        & & 2 & 
      12d* (4 NNs)  &
      5:1, 6:1 \\
    \hline
      FCC \newline 225 Fm$\overline{3}$m &
      \includegraphics[width=0.75in,valign=t]{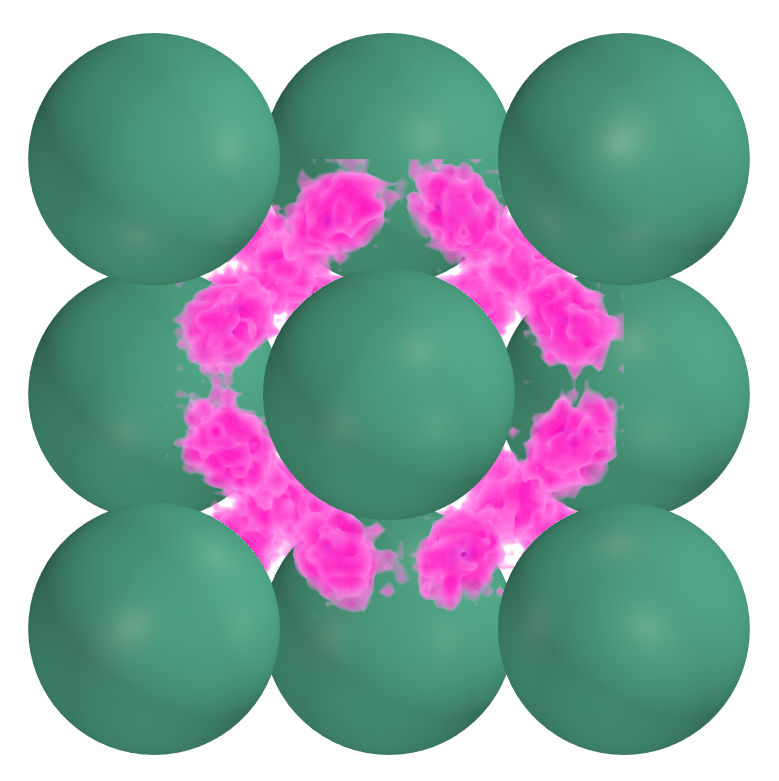} 
      \includegraphics[width=0.75in,valign=t]{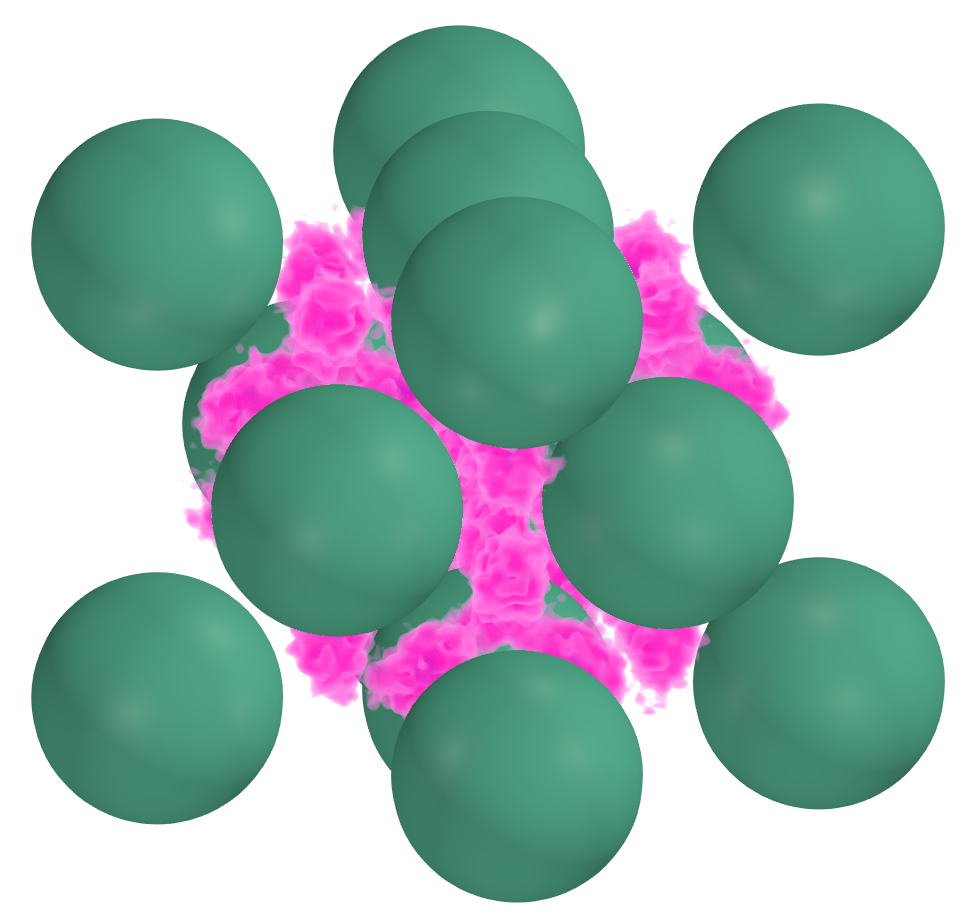} 
    & & 4 & 
      32f (3 NNs), \newline 8c* (3 NNs) &
      9:1, 10:1 \\
    \bottomrule
\end{tabular}
    \caption{\label{tab:lattice_layout} Observed lattices, defined by large particles at lattice points and small particles at interstitial sites, and arranged by the $n_s$:$n_l$ at which they are observed with a localized sublattice. Lower symmetry lattices appear in lower $n_s$:$n_l$ systems, and the $n_s$:$n_l$ ratio at which we observe a crystal type corresponds to: \#Wyckoff positions/\#lattice points, on a per-unit cell basis (for example: $12/2=6$ for a \gls{bcc}). In lower-symmetry lattices, small particles sit at Wyckoff positions with more nearest large particle neighbors (NNs) than those in higher-symmetry lattices. *12d positions in \gls{bcc} crystals and 8c positions in \gls{fcc} crystals are tetrahedral sites.}
\end{table}

\begin{figure}
    \includegraphics{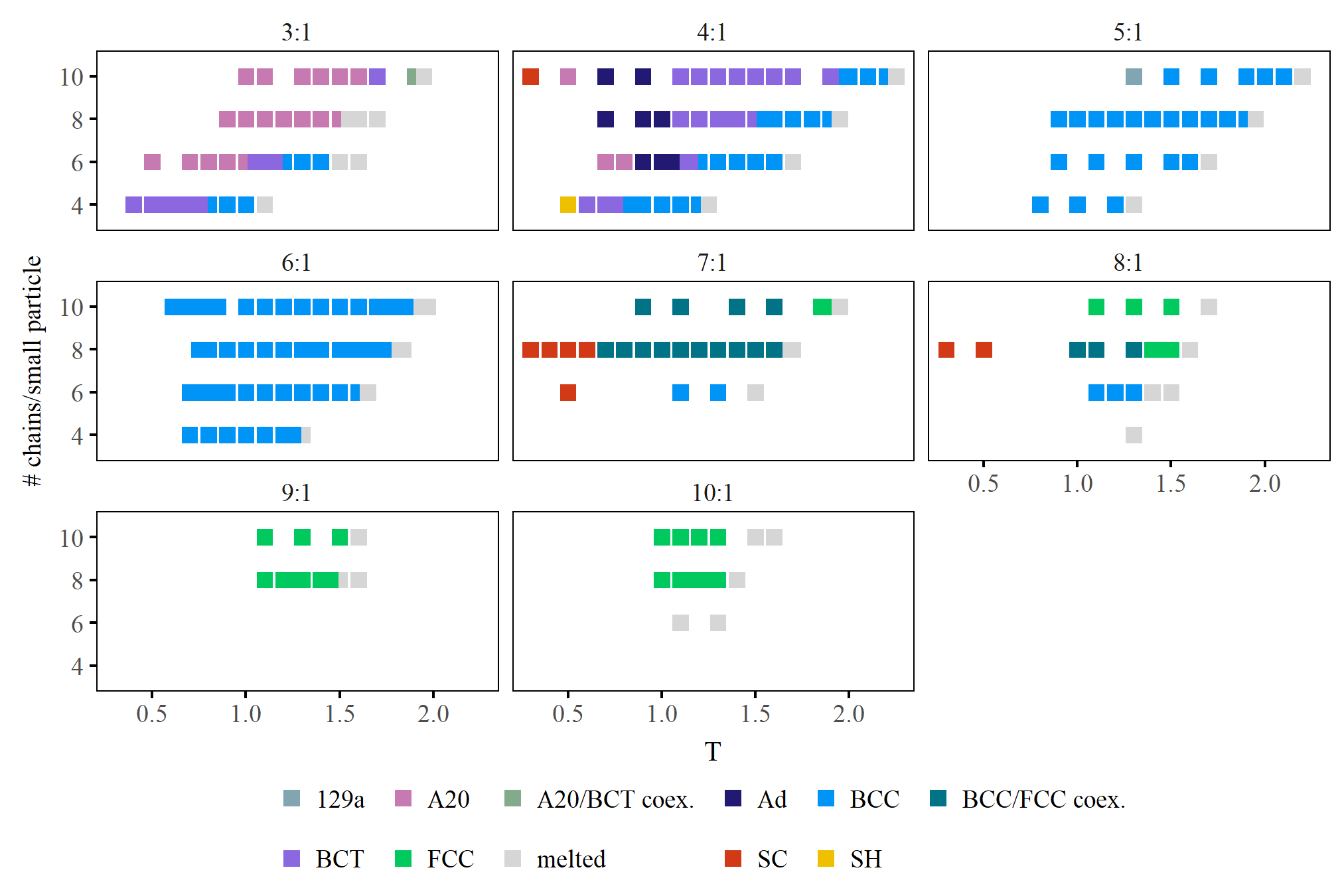}
    \caption{\label{fig:phase_diagrams} Phase diagrams of different $n_s$:$n_l$ values, as a function of reduced temperature \gls{T} (see \ref{sect:methods}) and number of chains per small particle. ``Coex." stands for ``coexistence" and 129a is an unknown crystal type in space group 129 defined in the Supplemental Material, Section II.A. The crystal structures observed only once are not analyzed in detail here. For the higher ratios where no data is shown for 4 or 6 chains, it is because no stable lattices were found. All data plotted in this paper is taken from simulations represented in these phase diagrams.}
\end{figure}

Table \ref{tab:lattice_layout} shows the most common crystals observed in our systems and the number ratios $n_s$:$n_l$ that produce them, and Figure \ref{fig:phase_diagrams} shows a phase diagram of all simulations studied in this work. The phase diagrams demonstrate visually that $n_s$:$n_l$ determines crystal structure, and the crystal properties in Table \ref{tab:lattice_layout} help explain trends present in the phase diagrams. For example, the value of $n_s$:$n_l$ at which a crystal structure is observed is consistent with the ratio of the number of lattice points (large particles) to the number of interstitial points associated with Wyckoff positions (small particles) of the unit cell. Explicitly, column 4 $=$ column 3 divided by column 2 (5:1 and 9:1 cases both contain interstitial vacancies). Note that some Wyckoff positions in A20 lattices are degenerate based on the observed lattice parameter ratios, so the A20 16h Wyckoff positions map onto 8 unique points instead of 16 and one 8g set maps onto 4 points. Table \ref{tab:lattice_layout} also demonstrates that the number of large particle nearest neighbors (NNs) interacting with each small particle decreases with increasing lattice symmetry and $n_s$:$n_l$ ratio. This is also an important approximation to the average potential energy interactions $U_\text{int.}$ between the two species. In summary, the findings demonstrated in Table \ref{tab:lattice_layout} and Fig. \ref{fig:phase_diagrams} show that decreasing $n_s$:$n_l$ results in lower-symmetry lattices with small particles sitting at lower-energy interstitial points. 
 
The most common lattices are A20, A$_\text{d}$, high symmetry \gls{bct}s, \gls{bcc}, and \gls{fcc}, though \gls{sh} and \gls{sc} are also observed. The non-cubic nature of \gls{bct}, A$_\text{d}$, and A20 requires a larger set of defining lattice parameters than the cubic crystals, and we observe multiple parameter ratios for each structure. For example, most \gls{bct} lattices with $n_s$:$n_l=$ 4:1 shown in Fig. \ref{fig:phase_diagrams} have the lattice parameter ratio $c/a = 2$. This is the configuration shown in Table \ref{tab:lattice_layout}, and it creates favorable conditions for 8 small particles in the unit cell, each of which interacts with 4 or 5 large particles depending on the site. 
However, some 3:1 and 4:1 \gls{bct} crystals in which the small particles have only 4 chains have $c/a=\sqrt{\frac{2}{3}}$ (not shown in Table \ref{tab:lattice_layout} for simplicity). We hypothesize that the interstitial sites in the more elongated \gls{bct} structure that allow for interactions with 5 large particle nearest neighbors require the small particles to have at least 5 chains. Therefore, small particles with only 4 grafted chains cannot stabilize those elongated structures. This is supported by a Fig. S3 in the Supplemental Material, which shows that small particles with 4 chains rarely interact with 5 large particles at once. Generally, \gls{bct} crystals only take discrete $c/a$ ratios corresponding to lattices of higher symmetry. For more details, see the Supplemental  Materials, Section II.A. and Section II.B.

A$_\text{d}$ lattices are also tetragonal and can be visually compared to \gls{bct} lattices in which an additional symmetry is broken because the conventional unit cell's central particle is not body-centered. The A$_\text{d}$ unit cell is defined by parameters $a$ and $c$ (similar to \gls{bct}) and $z$, which determines the offset of the central particles. When $z=0.5$, \gls{bct} symmetry is recovered. For all observed A$_\text{d}$ crystals $c/a =2$. However, there is a continuous increase of the $z$ parameter with \gls{T}, from $z\sim0.4$ at low \gls{T} to $z=0.5$ at the transition to \gls{bct} lattice with $c/a=2$. These local spatial changes as a function of temperature indicate the capacity for these colloidal crystals to be used as reconfigurable materials.
  
A20 crystals are orthorhombic and yet lower symmetry and more complex than the \gls{bct} or A$_\text{d}$ crystals. Their unit cells are defined by the ratios between $a$, $b$, and $c$, as well as a parameter $y$ that determines the lattice point placement within the unit cell. We observe two A20 crystal types with different lattice parameter ratios as a function of $n_s$:$n_l$. All 3:1 A20s have a consistent set of parameters $c/a$, $c/b$, and $y$, while the 4:1 A20 have another. Each parameter set results in different numbers of interstitial sites for the small particles. Additionally, due to the low symmetry of the A20 lattice, its parameters can be tuned to produce other lattices of higher symmetry. These include those observed at other values of $n_s$:$n_l$ and temperatures in this study, such as \gls{bcc} and \gls{fcc}. More details on all common lattices found in this study can be found in the Supplemental Material.

For almost all crystals listed in Table \ref{tab:lattice_layout}, a simple analysis of the potential energy landscape of a unit cell demonstrates why each lattice type is favorable at a given $n_s$:$n_l$ ratio. The landscapes were calculated with pairwise potentials between the large particles and one interactive chain bead, using the same method as described in \cite{Lopez-Rios2021}. The potential energy of a given point in a unit cell is the sum of the pairwise potential energy between a test particle (one interactive bead) located at that point within the unit cell and all large particles in the current and surrounding unit cells that contribute to the test particle's energy. This method only accounts for interactions between the large particle lattice and one interactive bead, and therefore does not take into account any small particle-small particle interactions or lattice vibration. However, even with these simplifications, the calculated energy landscapes can shed light on the spatial distribution of the particles. Each energy landscape shows potential energy wells (the most favorable locations for the interactive beads) and potential energy plateaus near zero (the least favorable locations for the interactive beads). For almost every lattice, the simulation results show that when the sublattice is localized, the interactive ends spend the most time in the energy wells, and the centers of the small particles spend the most time on the energy plateaus. This means we can predict the location of small particles once we know the unit cell of the large particle crystal, by identifying the location of the energy plateaus. The existence of these energy wells and their non-spherically symmetric distribution around the energy plateaus also highlights the importance of separation between the attractive component of the small particles and their cores, which in this case is due to the grafted chains. 

The fact that an analysis of a static energy landscape calculated with only small particle-large particle interactions can accurately identify the locations of the small particle centers indicates that the small particles do not substantially interfere with each other. A more detailed analysis of the \gls{bcc} case can be found in Lopez-Rios \textit{et al.} \cite{Lopez-Rios2021}, and a visual comparison between the energy landscape of a unit cell and the location of small particles can be found in the Supplemental  Material. There is one important exception: the FCC energy landscape shows plateaus at the octahedral and tetrahedral sites (Wyckoff positions 4b and 8c, respectively). However, we observe the small particles localizing at the 32f sites, where the energy plateaus are much smaller. In our systems that result in \gls{fcc} crystals, small particles never localize at the octahedral sites, and they localize at the tetrahedral sites only once the 32f sites are full (at ratios higher than $n_s$:$n_l=8$:1). We hypothesize that this is because the distance from the 32f sites to the large particles is shorter than the other sites which is needed to maintain a stable crystal with our system of short-range interactions. Additionally, there are fewer 4b and 8c sites in an \gls{fcc}, and for the $n_s$:$n_l$ ratio that would have filled those sites (3:1), there are more energetically favorable crystals available.

\begin{figure}
    \includegraphics{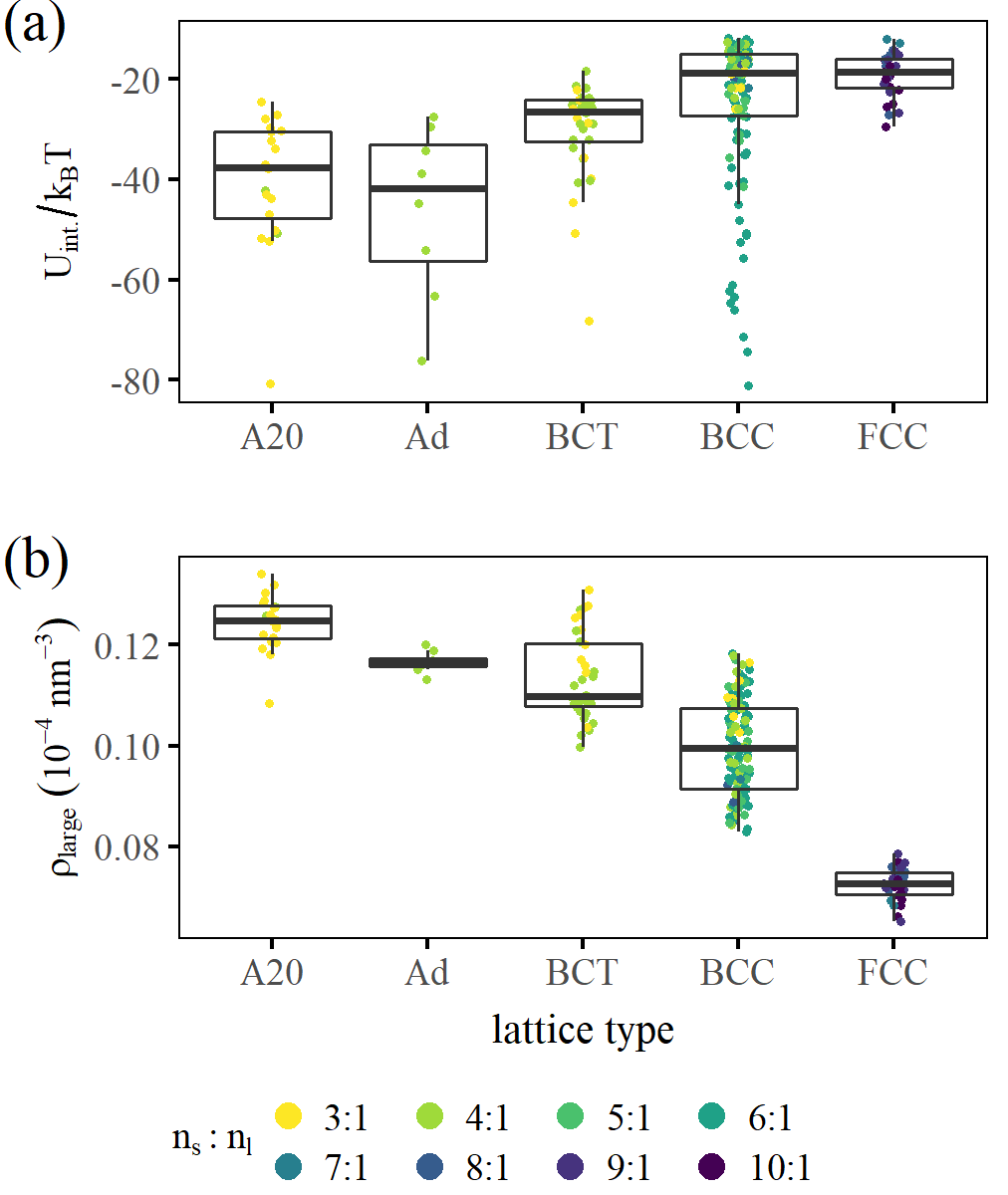}
    \caption{\label{fig:smallp_energy} (a) Average interaction energy $U_\text{int.}/k_BT$ per small particle, which quantifies the potential energy due to small particle-large particle interactions, and (b) number density of the large particles $\rho_\text{large}$ for each simulation that resulted in the most common crystals (A20, A$_\text{d}$, \gls{bct}, \gls{bcc}, and \gls{fcc}), arranged by crystal lattice type and colored by the value of $n_s$:$n_l$ to emphasize the effect of number ratio on lattice structure. Each data point represents a simulation under different conditions (temperature, number of chains, $n_s$:$n_l$), and the data shown comes from nearly all simulations in Fig. \ref{fig:phase_diagrams} that resulted in these common crystals. One very low temperature simulation with an A20 structure ($U_{int.}/k_BT < -100$) has been removed for clarity. Values of temperature and number of chains per small particle are not distinguished here.}
\end{figure}

Finally, as the number of small particles in the lattice increases (larger $n_s$:$n_l$ ratios), the energetic interaction between each small particle and the surrounding large particles becomes weaker and the packing density of large particles decreases. This can be seen in Fig. \ref{fig:smallp_energy}, which shows the average small particle-large particle interaction energy and system density for each of the common crystal lattices observed in our system. Almost all simulations shown in Fig. \ref{fig:phase_diagrams} are included. The number of large particles with which each small particle can interact decreases with increasing lattice symmetry; see the Supplemental  Materials for corresponding simulation data. For example, a \gls{bct} lattice with $c/a = 2$ has 8 interstitial sites, at which the small particles can interact with 4 or 5 large particles. Meanwhile, a \gls{bcc} unit cell contains 12 interstitial sites, and a small particle at any of those sites can interact with 4 large particles. Because \gls{bct} and \gls{bcc} unit cells each contain 2 lattice sites, the favorable sublattice sites are fully occupied at a 4:1 number ratio for a \gls{bct} and at 6:1 for a \gls{bcc}. If there are more small particles than can fit in the \gls{bct} interstitial sites, then the system's equilibrium lattice cannot be a \gls{bct} and it will instead form a \gls{bcc}. This pattern holds across all number ratios: systems with larger $n_s$:$n_l$ ratios form crystals containing interstitial sites that are greater in number but less energetically favorable. 

\subsection{Sublattice delocalization transition entropy and dependence on interstitial site filling}

We observe a transition to sublattice delocalization with increased \gls{T} for almost all assembled crystals. For some values of $n_s$:$n_l$, the transition to sublattice delocalization is a phase transition accompanied by a change in symmetry of the large particle lattice. For others, sublattice delocalization occurs as a smooth change rather than a phase transition. In the subsequent subsections, we detail the signatures of each observed transition behavior and corresponding lattice properties. 

For all values of $n_s$:$n_l$, we see two overarching trends. First, there is strong evidence that the transition to sublattice delocalization is driven by entropy. This is expected based on the form of the Gibbs free energy $\Delta G = \Delta H - T \Delta S$, the minimization of which determines the equilibrium crystal phase. $\Delta G$ is dominated by enthalpy $\Delta H$ at low \gls{T} and entropy $\Delta S$ at high \gls{T}. Entropic effects have also been experimentally shown to induce phase transitions of binary size-asymmetric colloidal crystals from energetically to entropically favored phases \cite{Bodnarchuk2010}. In our systems, we see this for all types of transition to sublattice localization. 

Second, increasing the chains per small particle increases the temperature at which the entropic transition occurs, effectively increasing the stability of the lattice. Crystal transition and melting temperatures increase approximately linearly with the number of chains per small particle for each value of $n_s$:$n_l$. Therefore, the addition of chains in most cases simply scales up the magnitude of the interaction between the large and small particles. There are a few exceptions to this rule, which will be discussed in following sections.

Note that the phenomenon of sublattice delocalization has been quantified using metallicity \cite{Girard2019} and occupied volume fraction \cite{Lopez-Rios2021}. However, these metrics are difficult to use for comparison between crystal phases due to convergence and normalization issues. We have previously found that sublattice delocalization is highly tied to small particle diffusion and lattice vibrations quantified as median lattice displacement \cite{Lopez-Rios2021}, both of which can be calculated more easily and are experimentally measurable. Therefore, we use these properties as measures of the degree of sublattice delocalization. 

\subsubsection{Phase transitions driven by lattice vibrations}

\begin{figure}
    \includegraphics{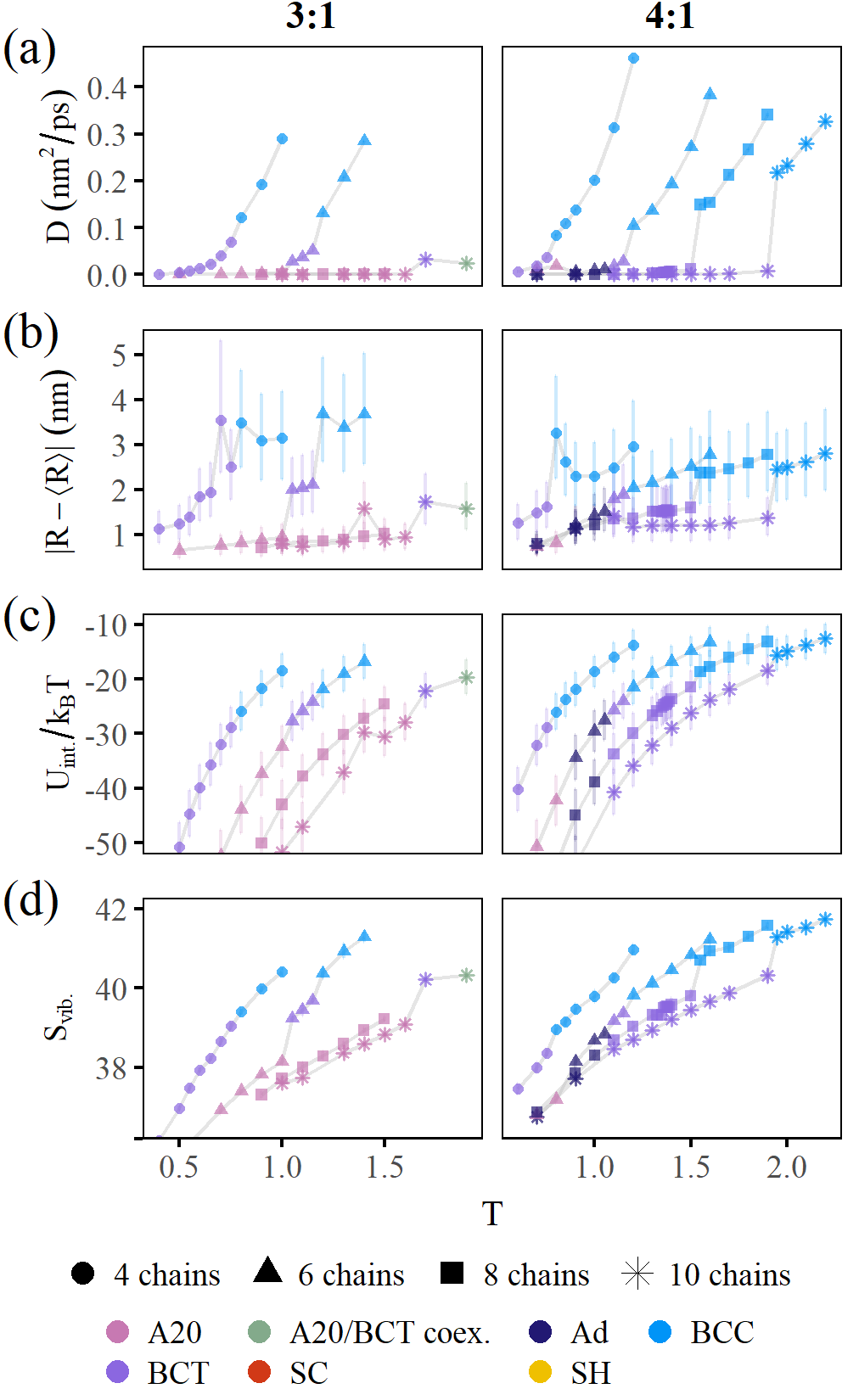}
    \caption{\label{fig:3and4} Lattice properties of 3:1 and 4:1 systems with 4, 6, 8, and 10 chains per small particle as a function of reduced temperature \gls{T} (see Section \ref{sect:methods}). All \gls{bct} crystals shown have lattice parameters $\frac{c}{a}=2$ except when the small particles have 4 grafted chains. (a) Diffusion constant, calculated as the slope of the mean squared displacement of the small particles in their linear (diffusive) regime. (b) Lattice fluctuations, quantified as the median displacement of large particles from their mean positions. (c) Average interaction energy $U_\text{int.}/k_BT$ per small particle. (d) Average lattice vibrational entropy of the large particles (as they occupy the crystal’s lattice points). All quantities show a jump around the phase transition to \gls{bcc} crystals. Some vary low-$T$ points not relevant to the transition have been removed for clarity.}
\end{figure}

For systems at low values of $n_s$:$n_l$, we observe a phase transition with increasing $T$ from a localized, low-symmetry lattice to a delocalized, higher-symmetry one, specifically \gls{bct} $\rightarrow$ \gls{bcc} and A20 $\rightarrow$ \gls{bct}. This is illustrated by a sharp increase in our two descriptors of sublattice delocalization, the diffusion coefficient (Fig. \ref{fig:3and4}(a)) and lattice vibrations (Fig. \ref{fig:3and4}(b)). The diffusion constant $D$ is calculated as the slope of the mean squared displacement of the small particles, which increases linearly at long time scales. Lattice vibrations $|R-\left<R\right>|$ are quantified as the median of the magnitude of the displacement of large particles from their mean positions. Both of these properties increase suddenly at the temperature of a crystal lattice transition, particularly a change to \gls{bcc}. It is also interesting to note that increasing the number of chains per small particle affects only the temperature at which this change occurs and does not impact the nature of the transition. That indicates that the addition of chains effectively increases the energetic interaction between the small and large particles, stabilizing the lattice against sublattice delocalization and melting. The exception to this is some systems with 4 grafted chains per small particle, which will be discussed later in this section.  

The observed transitions appear to be driven by entropy, and this is consistent with the observation that crystals lose energetic interactions while gaining entropy when transitioning to a \gls{bcc} with a delocalized sublattice. Fig. \ref{fig:3and4}(c) shows the average interaction energy per small particle in each system, as a function of $T$. As temperature increases, the interaction energy tends closer to zero, meaning that energetic interactions become weaker and less favorable. There is also a small jump at the transition to \gls{bcc} to weaker energetic interactions. This may occur for two reasons. First, the high-temperature \gls{bcc} lattice is generally less dense and therefore contains weaker interactions than the low-temperature \gls{bct} lattice. Additionally, all \gls{bct}s shown in Fig. \ref{fig:3and4} with more than 4 grafted chains per small particle have the lattice parameter ratio $\frac{c}{a}=2$. As indicated in Table \ref{tab:lattice_layout}, small particles interact with 4 or 5 neighboring large particles in this type of \gls{bct} crystal, but with only 4 in a \gls{bcc}, so some lose favorable interactions transitioning to a \gls{bcc}. Finally, delocalized small particles also occupy regions between interstitials which also decreases the number of interactions with neighboring large particles, as seen in Supplemental Material, Fig. S3. 

Vibrational entropy shows a similar signature. Fig. \ref{fig:3and4}(d) shows the lattice (large particle) vibrational entropy per large particle $S_\text{vib.}$ as a function of $T$, with clear jumps at the transition temperature. The vibrations of the large particles in a \gls{bct} with $\frac{c}{a}=2$ are more constrained parallel to the (001) planes due to denser packing in those planes. When the crystal transitions to a \gls{bcc}, the overall density of the system decreases and vibrations can be larger and more isotropic and contribute more to the entropy of the crystal (see the Supplemental Material, Section III.B., for details). Other forms of entropy are larger in the \gls{bcc} phase, as well. Delocalized small particles can occupy a larger volume than localized ones and therefore contribute to a larger entropy. Finally, \gls{bcc}s with $n_s$:$n_l =$ 3:1 or 4:1 contain an average of 6 and 4 interstitial vacancies per unit cell, respectively, and therefore their sublattices also have more configurational entropy as not all sublattice sites are filled. This is because, as indicated in Table \ref{tab:lattice_layout}, the sublattice of a \gls{bcc} is filled at $n_s$:$n_l = $ 6:1. However, having more interstitial vacancies should increase the lattice entropy, and the stability of the crystal will be negatively impacted as the melting temperature will be decreased.  

The nature of the transition can be further characterized by examining the behavior of the entropy of the system. Here, we consider $S_\text{vib.}$ to be representative of the total system entropy, as we know from previous work that lattice vibrations are highly tied to the other significant contributor to entropy, small particle delocalization, and it is more straightforward to calculate following  \cite{Pinsook2002}, see the Supplemental Material, Section III.F. A first order phase transition occurs at a discontinuity in the first derivative of the free energy, such as entropy. In Fig. \ref{fig:3and4}(d), there is a sharp jump in $S_\text{vib.}$ at the transition to sublattice delocalization when the number of grafted chains per small particle is greater than 4, strongly hinting at a discontinuity that would indicate the presence of a first order phase transition between a localized \gls{bct} and a delocalized \gls{bcc}. This is consistent with Landau  \textit{et al.} \cite{LandauVol5}, who state that a first order phase transition is expected between crystal phases when the curve of an appropriate order parameter connecting two phases of differing symmetry is not continuous. While the large particles of a \gls{bct} with $\frac{c}{a}=2$ can change continuously into a \gls{bcc}, this does not appear to be possible for the small particles, based on their interstitial positions. Therefore, it appears that the transition from \gls{bct} with $\frac{c}{a}=2$ and a localized sublattice to a \gls{bcc} with a delocalized sublattice is first order. Additionally, estimates of the specific heat capacity corroborate these conclusions and are given in the Supplemental Material, Section III.I. 

To further confirm the nature of this transition, we look to the phonon-driven \gls{mit} in vanadium dioxide (VO$_2$). The sudden change from an insulating to a conducting state in VO$_2$ as a function of \gls{T} is enabled by a phase transition to a more symmetric and entropic crystal phase, in which a strong metallic electron-phonon correlation was detected consistent with a Peierls \gls{mit} \cite{Budai2014}. Budai \textit{et al.} identified the electron-phonon correlations using the phonon density of states, which narrows towards lower vibrational frequencies in the metallic phase, and anharmonic vibrational modes impeding the filling of lower energy orbitals only in the metallic phase. In our systems that appear to exhibit a first-order sublattice transition, we also find a bias towards lower vibrational modes in crystals with a delocalized sublattice. There is also evidence of anharmonic modes due the expanding lattice parameter of the metallic \gls{bcc} crystals as a function of temperature. Finally, we calculate a greater momentum exchange in crystals with a delocalized sublattice, which is most likely due to small particles being more homogeneously distributed throughout the crystal. See the Supplemental  Materials for the vibrational density of states (following Dickey \textit{et al.} \cite{Dickey1969}) and the momentum cross-correlation (following Verdaguer \textit{et al.} and Ishida \cite{Verdaguer1998,Verdaguer2001,Ishida2011}) for the case of a system that exhibits a first-order sublattice transition.    

The exception to this discussion is the cases in which the small particles have 4 grafted chains. In those cases, the entropy in Fig. \ref{fig:3and4}(d) appears to be continuous but with a change in slope at the transition, indicating a discontinuity in the specific heat capacity, rather than entropy. According to Landau \textit{et al.} \cite{LandauVol5}, a discontinuity in the specific heat is to be expected for continuous phase transitions, specifically between crystal types than can continuously change into one another. While we would need more data to confidently determine the classification of this phase transition, it is also consistent with our intuition that the phase transition for 4 grafted chains per small particle be continuous. This is because the low temperature \gls{bct} crystals have $\frac{c}{a}=\sqrt{\frac{2}{3}}$ when the small particles have only 4 grafted chains. As discussed in Section \ref{sect:compositions}, we believe that small particles with only 4 grafted chains cannot stabilize a \gls{bct} with $\frac{c}{a}=2$. However, for \gls{bct} with $\frac{c}{a}=\sqrt{\frac{2}{3}}$, the interstitial sites appear to be such that it is possible for both the small and large particles to continuously change to their \gls{bcc} lattice sites. Note that the 3:1 system with 4 chains per small particle also transitions through an unclassified \gls{bct}; see Fig. S1 in the Supplemental Material for more information.

Finally, other low-temperature transitions between crystal types are shown in Fig. \ref{fig:3and4}, for example A20 $\rightarrow$ \gls{bct} and A$_\text{d}$ $\rightarrow$ \gls{bct}. These transitions exhibit interesting changes in symmetry; however, we do not study those changes here because they are not accompanied by a change in sublattice delocalization. 

\subsubsection{Smooth change to sublattice delocalization driven by stoichiometry}

\begin{figure}
    \includegraphics{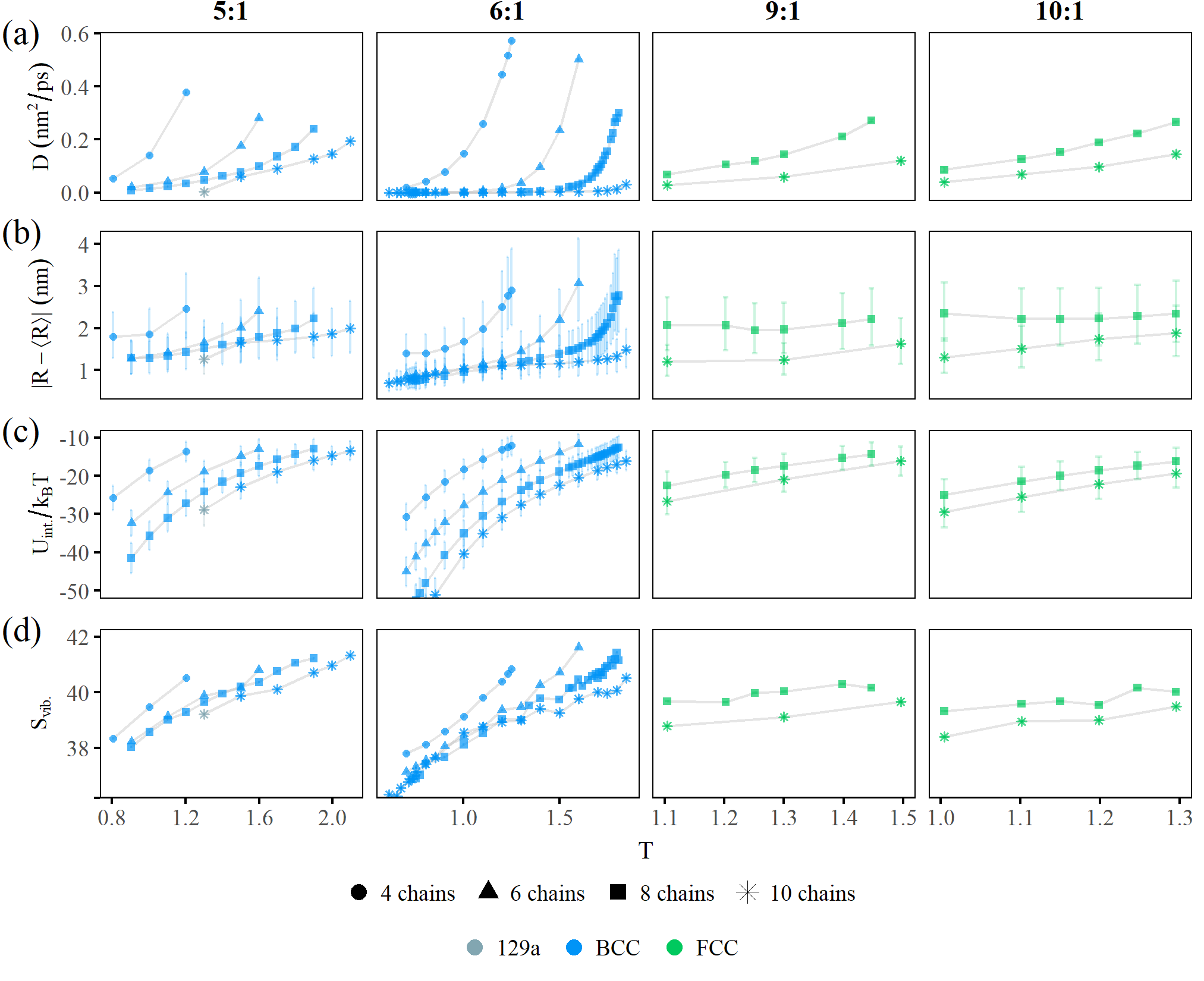}
    \caption{\label{fig:56and910} Lattice properties of 5:1, 6:1, 9:1, and 10:1 systems with 4, 6, 8, and 10 chains per small particle as a function of reduced temperature \gls{T} (see \ref{sect:methods}). Data from the 6:1 system is taken from \cite{Lopez-Rios2021} and included for comparison. (a) Diffusion constant. (b) Lattice fluctuations. (c) Average interaction energy $U_\text{int.}/k_BT$ per small particle. (d) Average lattice vibrational entropy of the large particles. All show a smooth increase in diffusion and lattice vibrations, indicating a change to delocalization similar to that explored in the 6:1 system.}
\end{figure}

At $n_s$:$n_l$ near the stoichiometric values for \gls{bcc} crystals (6:1) or \gls{fcc} crystals (10:1), the transition to delocalization of the small particles is gradual and not a true phase transition. In these cases, the sublattice delocalizes slowly over a range of temperatures and the large particle lattice never changes structure. This can be seen in Fig. \ref{fig:56and910}. Note that, again, as \gls{T} increases, diffusion and vibrational entropy of the large particles increase at the expense of the magnitude of the interaction energy. We hypothesize that this is because the \gls{bcc} and \gls{fcc} lattices are the most symmetric and stable crystals available to systems at lower and higher $n_s$:$n_l$ ratios, respectively. Specifically, \gls{bcc} lattices are entropically stabilized at high \gls{T} \cite{Alexander1978,Sprakel2017}, so we do not expect a \gls{bcc} to transition to another crystal with increasing \gls{T} as long as the number of small particles does not exceed the number of interstitial sites (\textit{i.e.} a number ratio greater than 6:1). At higher number ratios, which would otherwise result in \gls{bcc} lattices with interstitial defects, \gls{fcc} crystals are stable simply based on stoichiometry. This will be discussed further in the next subsection.  

The 6:1 system is an exemplar of this behavior and has been studied in detail by Lopez-Rios \textit{et al.} \cite{Lopez-Rios2021}. The conclusions of that study were that lattice vibrations and sublattice delocalization are strongly tied, and the temperature of the onset of both is dependent on the number of chains per small particle. We have found this to be true in general for systems that do not exhibit a lattice transition with temperature.

\subsubsection{Phase transition driven by interstitial defects}
\label{sect:twophase}

\begin{figure}
    \includegraphics{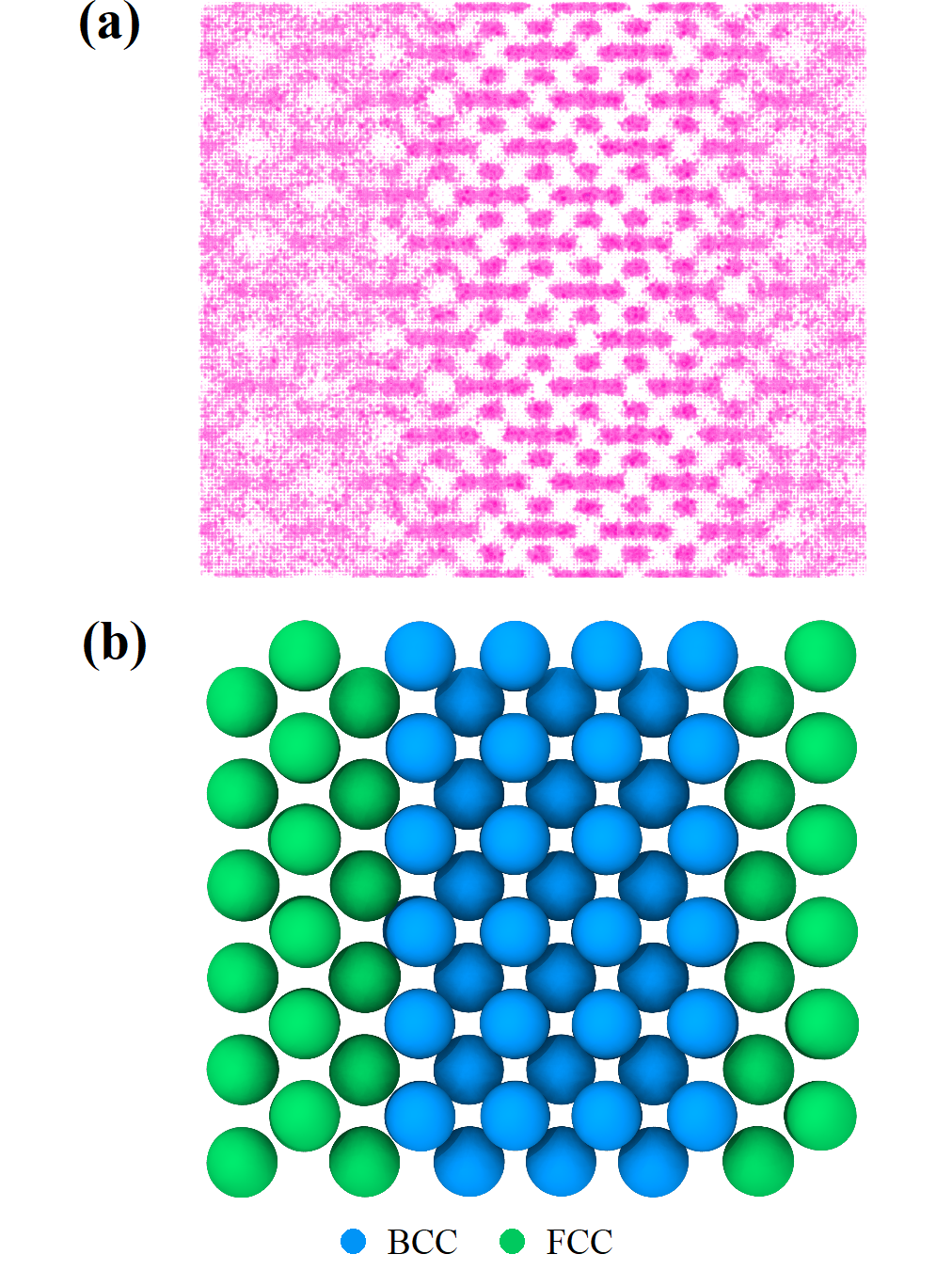}
    \caption{\label{fig:mixture_image} \gls{bcc}/\gls{fcc} coexistence in a simulation with $\gls{T}=1.6$, 8 chains per small particle, and $n_s$:$n_l$ = 7:1. Stable localized \gls{bcc} and delocalized \gls{fcc} portions can be seen in (a) a snapshot of the locations of the small particle centers and (b) the averaged positions of the large particles, colored by crystal phase.}
\end{figure}

\begin{figure}
    \includegraphics{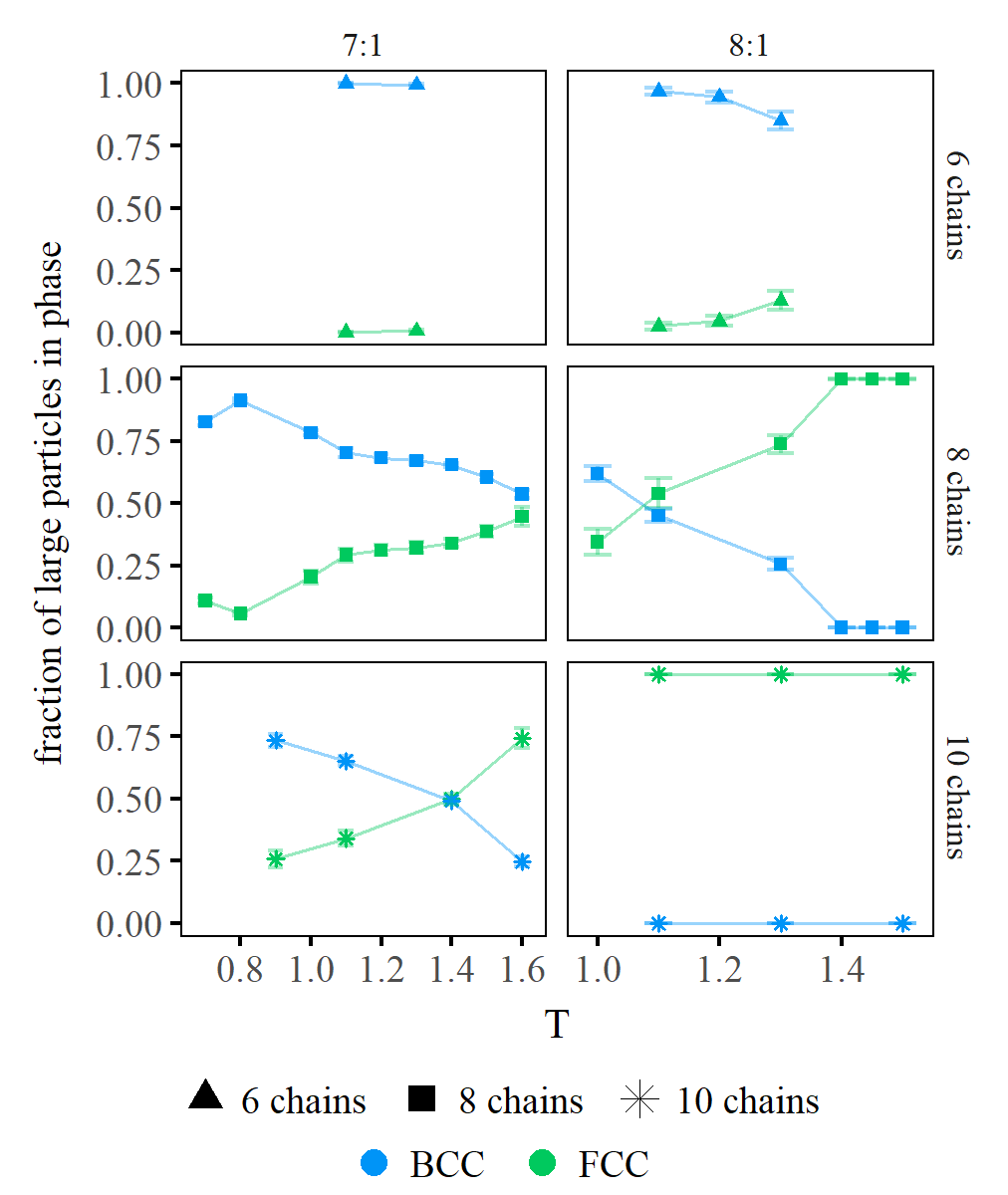}
    \caption{\label{fig:7and8_composition} Fraction of large particles in the simulation in the \gls{bcc} and \gls{fcc} phases for 7:1 and 8:1 systems. The small portion of particles in neither phase is not shown. Increasing both reduced temperature \gls{T} and number of chains per small particle increases the percentage of the delocalized \gls{fcc} lattice. These compositions were tested for stability with annealing techniques and at multiple system sizes.}
\end{figure}

For systems with $n_s$:$n_l =7$:1 and 8:1, between the stoichiometric number ratios for \gls{bcc} and \gls{fcc}, we observe a stable two-phase coexistence between a localized \gls{bcc} and delocalized \gls{fcc}. Coexistence is an indication of a first-order transition between the two phases, and an example is in Fig. \ref{fig:mixture_image}. Experimental evidence of a \gls{bcc}/\gls{fcc} mixture in colloidal crystals was reported at a small particle-large particle number ratio between those required for fully \gls{bcc} or fully \gls{fcc} crystal structures \cite{Girard2019}.

\gls{fcc} lattices in these systems appear only at high number ratios (7:1, 8:1, 9:1, 10:1), as can be seen in Table \ref{tab:lattice_layout}. This is also consistent with Girard \textit{et al.} \cite{Girard2019}, who observed FCC lattices when the concentration of small particles in solution was high. In our 7:1 and 8:1 systems, the \gls{fcc} phase appears to be the result of interstitial defect attraction. It has been established that \gls{bcc} lattices with small particles localized at the usual tetrahedral sites ($n_s$:$n_l=6$:1) are stable. At a $n_s$:$n_l$ of 7:1 or 8:1, however, a fully \gls{bcc} system would contain 2-4 interstitial defects per unit cell, which is energetically unfavorable. As has been demonstrated by van der Meer \textit{et al.}  \cite{VanDerMeer2017}, interstitial defects in colloidal systems show long-range attraction. Therefore, the defects in the \gls{bcc} system gather when there are strong small particle-large particle interactions (8 and 10 chains per small particle). At very low temperatures, they collect at a grain boundary; a snapshot of this is shown in the Supplemental Material, Fig. S8. At moderate and high temperatures, they collect and expand the lattice, resulting in a \gls{fcc} phase with a delocalized sublattice coexisting with the \gls{bcc} phase with a localized sublattice. This is consistent with Fig. \ref{fig:7and8_composition}, which shows that 8:1 systems have a higher \gls{fcc} fraction at a given \gls{T} and number of chains per small particle. 

As can also be seen in Fig. \ref{fig:7and8_composition}, increasing \gls{T} results in an increased fraction of the system in the \gls{fcc} phase. This indicates that the transition between a localized \gls{bcc} and delocalized \gls{fcc} is at least in part driven by entropy. Each small particle interacts with 4 large particles in a \gls{bcc} lattice when localized and only 3 in an \gls{fcc} lattice (and even fewer when delocalized due to spending less time at energetically favorable sites). Therefore, the transition from the \gls{bcc} phase to the \gls{fcc} phase results in an energy penalty, which is compensated for by a gain in entropy in the form of small particle mobility and lattice vibrations in the \gls{fcc} phase. 

Lastly, increasing the number of chains per small particle results in a higher \gls{fcc} fraction, which deviates from the general rule that adding chains simply increases lattice stability. We hypothesize that this is due to the difference in the unit cell energy landscape between the \gls{bcc} and \gls{fcc} lattices. The energy landscape of the \gls{fcc} is overall shallower and more homogeneous than that of the \gls{bcc}, as there is little overlap between the attractive regions around the large particles (see the Supplemental Material, Table S1, for comparisons). In contrast, the \gls{bcc} unit cell energy wells are deep and localized in spaces between large particles. 
Therefore, it may be that small particles interact favorably only with \gls{fcc} energy landscapes when there are more chains and when those chains are configured more isotropically. 
This may explain why size-asymmetric binary colloidal systems composed of spherical particles have only seen \gls{fcc} lattices \cite{Filion2011,Lin2020} and why other crystals such as \gls{bcc} have been observed only with the existence of flexible chains on the small particles \cite{Lopez-Rios2021,Girard2019}.

\subsection{Stability as a function of number ratio $n_s$:$n_l$}

\begin{figure}
    \includegraphics{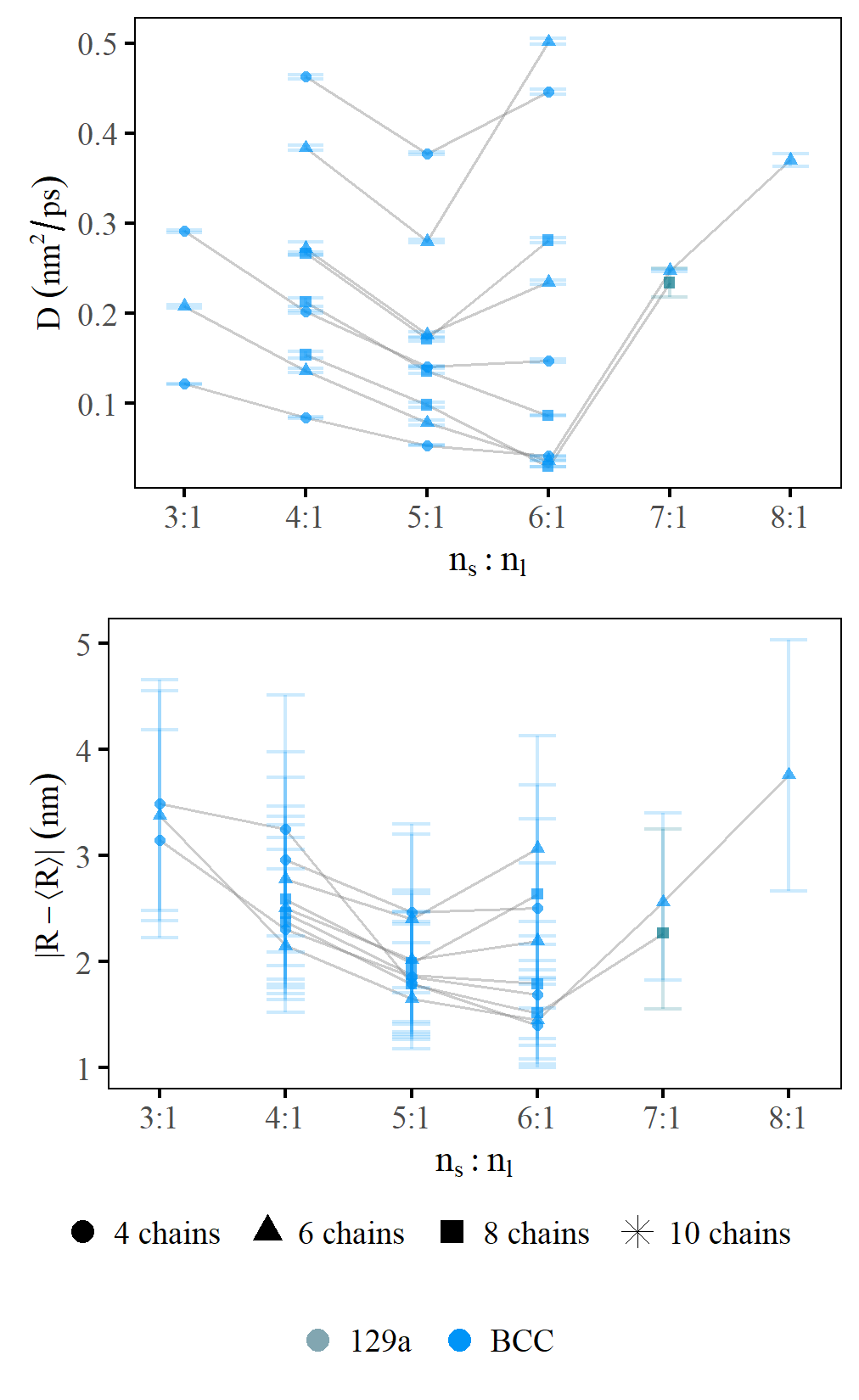}
    \caption{\label{fig:by_EE_true} (a) Diffusion coefficients $D$ and (b) lattice vibrations as a function of $n_s$:$n_l$ for simulation groups that do not exhibit a phase transition. Both $D$ and lattice vibrations both show a minimum at 6:1, similar to the metallicity found by Girard \textit{et al.} \cite{Girard2019}. Lines connect points with the same value of \gls{T} and number of chains, and lines are not drawn between non-adjacent points, or if any number of chain-\gls{T} combination has fewer than 3 data points. Though it is not visually depicted, higher values of $D$ and lattice vibrations for a given number of chains correspond to higher temperatures.}
\end{figure}

\begin{figure}
    \includegraphics{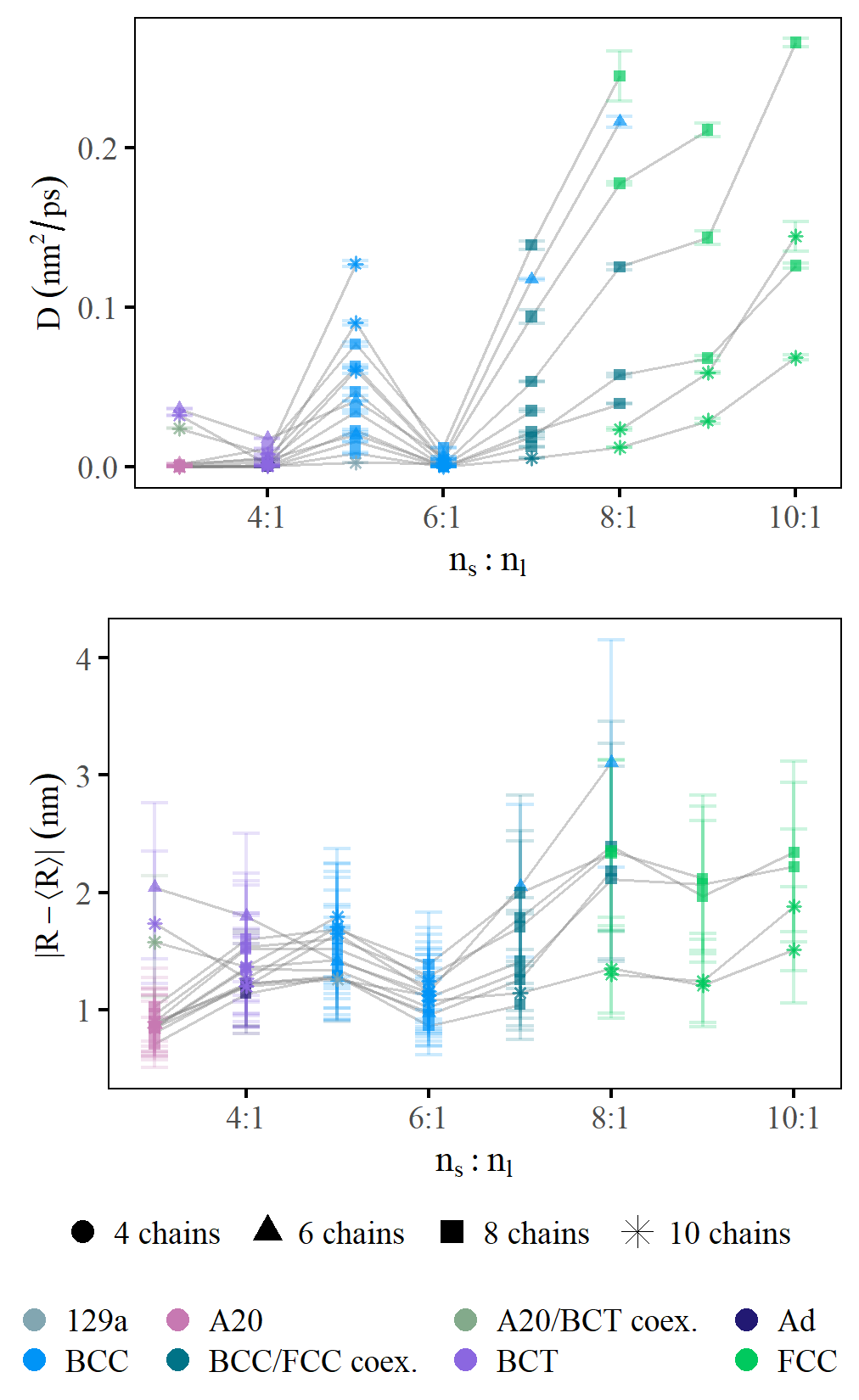}
    \caption{\label{fig:by_EE_false} (a) Diffusion coefficients $D$ and (b) lattice vibrations as a function of $n_s$:$n_l$ for simulation groups that do exhibit a phase transition. Both quantities show minima at number ratios corresponding to compound values for \gls{bcc} and A20 crystals. Lines connect points with the same value of \gls{T} and number of chains, and lines are not drawn between non-adjacent points, or if any number of chain-\gls{T} combination has fewer than 3 data points. Though it is not visually depicted, higher values of $D$ and lattice vibrations for a given number of chains correspond to higher temperatures.}
\end{figure}

Overall, crystals are more stable and have lower sublattice delocalization when small particles saturate their interstitial sites. This is highlighted in Fig. \ref{fig:by_EE_true} and Fig. \ref{fig:by_EE_false}, which show diffusion and lattice vibrations as a function of $n_s$:$n_l$ for systems with different \gls{T}-chain number combinations. For clarity, data is separated by whether there is a crystal phase transition as a function of $n_s$:$n_l$. A minimum in both quantities appears at 3:1, 4:1, and 6:1 (for the 3:1, 4:1 and 6:1 systems that form \gls{bcc}, A20 and \gls{bct} lattices with a fully saturated sublattice). Meanwhile, the 5:1 (\gls{bcc} crystals) and 9:1 (\gls{fcc} crystals) ratios both contain inherent vacancies that diffuse, since \gls{bcc} and \gls{fcc} interstitials are fully occupied at 6:1 and 10:1 ratios, respectively. Additionally, according to Table \ref{tab:lattice_layout}, \gls{fcc} lattices and their interstitials are less tightly bound than in \gls{bcc} lattices and therefore should show more delocalization at a given \gls{T}. It is not included, but lattice vibrations also show minima at 3:1 and 6:1 ratios. 

The predominant appearance of \gls{bcc} lattices over the entire phase space explored may be due to their stabilization by entropy \cite{Sprakel2017}. Their lattice vibrations are isotropic and this garners them additional structural stability as a function of temperature that enables a larger degree of sublattice delocalization than other lattices. For similar reasons, \gls{bcc} lattices have been suggested as optimal superionic conductors in atomic systems \cite{Wang2015}.

\section{Discussion and conclusions}
\label{sect:discussion}

In summary, highly size-asymmetric binary colloids assemble into a variety of crystals that exhibit varying levels of sublattice delocalization. For temperatures at which the sublattice is localized, the crystal structure is determined by energetic interactions between the small and large particles. Crystals with a lower number ratio $n_s$:$n_l$ form lower-symmetry crystals whose unit cell potential energy landscapes contain many deep wells. As $n_s$:$n_l$ increases, crystals become more symmetric and the wells become shallower. As a function of \gls{T}, we observe different types of entropically driven transitions to sublattice delocalization. In some cases, this transition occurs along with a symmetry change of the large particles, always from a lower-symmetry lattice to a higher-symmetry lattice containing more interstitial vacancies. In others, when the lattice is in a cubic configuration (these are entropically stabilized) or already contains inherent vacancies, there is not a phase transition to sublattice delocalization but rather a smooth change.

Additionally, we observe the appearance of different crystal lattices as a function of $n_s$:$n_l$ at constant \gls{T}. This is consistent with experiments using DNA functionalized \gls{nps} \cite{Girard2019,Cheng2021} even though hybridization DNA chemistry employed in those studies complicates experiments by including the presence of non-hybridized DNA chains that could act like depletant particles \cite{Girard2019,Cheng2021}. In particular, the transition we found from \gls{bct} to \gls{bcc} as $n_s$:$n_l$ increases agrees with Fig. 3 of Cheng \textit{et al.} \cite{Cheng2021}; note that in \cite{Cheng2021}, ``valency" is  the number of linkers per small particle and not the number ratio of small (``electron equivalent”) particles to large particles as it was defined in \cite{Girard2018} and in Fig. S29 in the SI of \cite{Girard2019}.

We report minima as a function of lattice vibrations and the diffusion constant of the small particles as a function of $n_s$:$n_l$ in Fig. \ref{fig:by_EE_false} and Fig. \ref{fig:by_EE_true}. It is tempting to compare these minima to the minima in metallicity identified by Girard and Olvera de la Cruz \cite{Girard2019}, which were found for each crystal phase (\gls{bcc}, \gls{fcc}, and Frank-Kasper A15) and which correspond to the compound value of $n_s$:$n_l$ for that phase. The behavior of the lattice vibrations and diffusion constants is similar, indicating that these reflect the same underlying phenomenon. However, we found that it is difficult to compare metallicity values between phases due to normalization and numerical convergence issues; using the more physically measurable values resolves these problems. Plotting indicators of sublattice delocalization in multiple phases on the same axis allows us two additional insights. First, this enables us to compare behavior between phases. We find that there are still minima at the saturation values for some lattices (A20, \gls{bct}, and \gls{bcc}), but that the minimum for \gls{fcc} found in \cite{Girard2019} does not appear because competition between \gls{bcc} and \gls{fcc} phases allows for a coexistence not seen in \cite{Girard2019}. Second, we see that the studied assemblies are generally more stable in the form of a \gls{bcc} lattice, whether their sublattice is localized or delocalized. Most of the low-symmetry crystal phases transition to \gls{bcc} at high temperatures, and \gls{bcc} only fully transitions to \gls{fcc} when the number of interstitial defects is very high. \gls{bcc}’s greater structural stability is consistent with observations that \gls{bcc} crystals are entropically stabilized near their melting point in colloidal assemblies \cite{Sprakel2017} (even without a sublattice). For these systems, the result of \gls{bcc} lattice stability is that these crystals can maintain a delocalized sublattice for a wider range of temperatures than other crystals. Additionally, Wang \textit{et al.} predicted that superionic materials with a \gls{bcc} structure should exhibit the highest conductivity \cite{Wang2015}, which is of particular interest for applications in solid-state batteries. Our results agree with this for the case of \gls{nps} and confirm the stability of \gls{bcc} colloidal crystals with delocalized sublattices.

It is intriguing to find similar behavior at multiple length scales, from sublattice melting in superionic materials to the \glsentryfull{mit} in inorganic materials to sublattice delocalization in colloidal binary crystals. Although colloidal systems are more flexible and tunable due to the lack of any sort of charge neutrality constraint on composition, they exhibit similarities to superionic materials in both structure and dependence on lattice vibrations, explored previously by Lopez-Rios, \textit{et al.} \cite{Lopez-Rios2021}. There are also structural and delocalization transition analogs between colloidal crystals and materials exhibiting an \gls{mit}. For example, at low \gls{T} and $4$:$1$ number ratio, crystal phases resemble the actinide crystal structures, where increasing the number of chains per small particle is analogous to increasing the atomic number. Systems with 4, 6, and 8 chains per small particle assemble into \gls{bct} ($c/a=\sqrt{2/3}$), A20, and A$_\text{d}$ lattices, which have the same symmetry as protactinium, $\alpha$-uranium, and $\beta$-neptunium, respectively. Increasing \gls{T} of these and other systems, we observe a transition to sublattice delocalization strongly driven by lattice vibrations. When accompanied by a change of lattice symmetry, this resembles a Peierls \gls{mit}, a transition driven by strong correlations between phonons and metallic electrons. For colloidal crystals, this can be thought of as a continuous pumping of momentum of the vibrating large particles to the diffusing small particles. As crystals become more symmetric, lower vibrational frequencies are available, which prolongs the exchange of momentum between the two species given their large vibrational wavelengths. Such tunability as a function of \gls{T} makes these colloidal crystals possible candidates for exploration as colloidal photonic crystals \cite{Cai2021, He2020}. There are other types of \gls{mit}, such as the Mott \gls{mit}, which is driven by the interactions and correlations between the smaller species. We observe stronger sublattice localization as a function of $n_s$:$n_l$ with a greater number of grafted chains, which is similar to the behavior of metallicity \cite{Girard2018}. This may be seen as a Mott-like transition, where the delocalized lattice may be suppressed by the addition of grafted chains on the small particles as was alluded by Girard \textit{et al.} \cite{Girard2018}. However, in some cases, the addition of grafted chains may also change the crystal lattice structure, which complicates this analogy.  

There is still more to explore. It is possible that by including the deformability of the large particles, one might increase the range of accessible phases such as the Frank Kasper A15 phase \cite{Girard2019}. Furthermore, given that lattice vibrations drive the transition to sublattice delocalization and between host lattices, it would be interesting to consider how impinging acoustic waves or acoustic shock waves would affect the properties of these colloidal crystals for further applications.

\section{Simulation methods}
\label{sect:methods}

In the model, as described in Fig. \ref{fig:model} (and also in \cite{Lopez-Rios2021}), we change the temperature \gls{T}, the number of chains per small particle, and the small particle-large particle number ratio $n_s$:$n_l$. Temperature \gls{T} is expressed in reduced units, such that $T = \frac{k_B T'}{\varepsilon}$, where $T'$ is the input temperature and $\varepsilon$ is the energy unit of the simulation, in our case $T = 1 = 5/3$ kJ/mol.

All simulations were conducted at constant number of particles $N$, temperature $T$, and pressure $P$. The pressure $P$ was the same in all simulations, $P$ = 2 Pa (approximately 2\% of atmospheric pressure). Simulations at low $P$ simplify the possible contributions to the formation and stability of a crystal such that only two terms remain, energetic and entropic. The pair potential interactions within our model arise from an attractive Gaussian potential between large particles and the termini of the grafted chains $U_{\mathrm{Gaussian}}(r)$ (Eq. \ref{eq:gauss}), as well as excluded volume interactions amongst all particles, modeled using the Weeks-Chandler-Andersen (WCA) potential $U_{\mathrm{WCA}}(r)$ (Eq. \ref{eq:wca}). The grafted chains are bonded with harmonic potentials, and no angle or dihedral potential is employed. We also used the HOOMD-blue \texttt{xplor} option which prevents artificial discontinuities in $U_\text{Gaussian}(r)$ as it decays to zero. Parameters used are shown in Table \ref{subtab:params-fix}. 
\begin{align}
    \label{eq:gauss}
    U_\text{Gauss} (r)   
        =&
        -\varepsilon
        e^{-\frac{1}{2} \left( \frac{r}{\sigma_\text{gauss}} \right)^2}
        &&\text{ for } r \le r_\text{cutoff} \\
    \label{eq:wca}
    U_\text{WCA} (r)
        =& ~
        4 
        \left( 
            \left( \frac{\sigma}{r} \right)^{12}
            -
            \left( \frac{\sigma}{r} \right)^{6}
        \right)
        -
        4 
        \left( 
            \left( \frac{\sigma}{2^{1/6} \sigma} \right)^{12}
            -
            \left( \frac{\sigma}{2^{1/6} \sigma} \right)^{6}
        \right)
        &&\text{ for } r \le 2^{1/6} \sigma
\end{align}
where $\sigma = R_A + R_B$ is the sum of the radii of the interacting species. 

\begin{table}
    \centering
    \label{subtab:params-fix}
        \begin{tabular}{|cc|}
            \hline 
            \textbf{Parameter} & \textbf{Value} \\
            \hline 
            $R_\text{large particle}$ & 10.5 nm \\
            $R_\text{small particle center}$ & 1.0 nm \\
            $R_\text{chain bead}$ & 1.0 nm \\
            $R_\text{interactive chain end bead}$ & 0.5 nm \\
            $\varepsilon$ & 70 kJ/mol \\
            $\sigma_{gauss}$ & 4.8 nm \\
            $R_\text{cutoff}$ & 8.4 nm \\
            \# non-interactive beads/chain & 3 \\
            % large particle mass & 3 R  \\
            % small particle mass & 3 R \\
            % chain bead mass & 1 R \\
            \hline
        \end{tabular}
    \caption{ \label{tab:params}
    Parameters used in the present study.   
    }
\end{table}

All simulations were initiated with 6$\times$6$\times$6 unit cells in the simulation box with either an \gls{fcc} or \gls{bcc} lattice with lattice parameter $a = 60$ nm. They were all energetically and thermally equilibrated using NVE integration and later Langevin integration, respectively, then depressurized to their final pressure. This sequence lasted 312 ns. Finally, the simulations were run at their final pressure $P$ = 2 Pa for at least 8.44 $\mu$s, the first 1.38 $\mu$s of which was considered an equilibration period and not used for analysis. Simulation code is available upon request.

To determine the crystal phase resulting from a simulation of a given set of parameters (\gls{T}, $n_s$:$n_l$, number of grafted chains, and initial configuration), we analyzed the pair correlation function ($g(r)$) of the large particles. See the Supplemental Material, Section I.A. for details.  

While exploring parameter space by changing $n_s$:$n_l$, it is important to ensure that the crystal configurations we are reporting are equilibrium configurations. To that end, we initialized many $n_s$:$n_l$-\gls{T}-chain parameter combinations in multiple ways, \textit{i.e.} \gls{bcc} and \gls{fcc} with an unphysically large lattice parameter, about 3 -- 5 times any lattice parameter from an equilibrated lattice structure of this study. If both simulations equilibrated to the same crystal configuration, we considered that configuration to be the lowest free energy state and selected only one to include for analysis in our final set. If the simulations had different results, we annealed both using various techniques described in the Supplemental Material, Section I.A. until both equilibrated to the same configuration. Note that a simulation initialized as an \gls{fcc} has twice as many particles as one initialized as a \gls{bcc} (because the \gls{fcc} unit cell contains twice as many particles), so this procedure of different initialization is also a test for finite size effects. 

Sometimes, this annealing process resulted in one version of the simulation with a bulk monocrystal and another in a polycrystal with grain boundaries. It has been observed experimentally that annealing polycrystalline colloids does not always results in a monocrystalline phase, possibly because of the similarity between the melting temperature and the temperature required to remove  grain boundaries (see the Supplementary Discussion of \cite{Auyeung2013}). If, after a few rounds of annealing, the two did not converge to exactly the same configuration, we chose to use the simulation resulting in the monocrystal. This is because polycrystals are always higher energy than monocrystals, and the purpose of the current study is the understand bulk crystals based on different parameter sets. Including polycrystals and the added complexity of grain boundaries is outside of these bounds. 

Finally, finite-size effects are often associated with seeing two-phase coexistence in an NPT simulation. To test whether simulation size played a role in the existence of two phase in our 7:1 and 8:1 systems, we ran and annealed all points of 7:1 and 8:1 systems in at least two initial configurations (usually \gls{bcc} and \gls{fcc}). Simulations of different sizes resulted in very similar \gls{bcc} to \gls{fcc} ratios, which are shown in Fig. \ref{fig:7and8_composition}. We tested one system (7:1, 8 chains per small particle, $T*=1.3$) with 432, 864, and 2000 large particles and saw roughly the same \gls{bcc} to \gls{fcc} ratio in all three simulations. 

System topology for the simulation was built using Hoobas \cite{Girard2019a}. Simulations were run with Hoomd-blue \cite{Anderson2020,Anderson2008} and analyzed using MDAnalysis \cite{Gowers2016,Michaud-Agrawal2011}. Images were created with Mayavi \cite{Ramachandran2011} (Fig. \ref{tab:lattice_layout} and \ref{fig:mixture_image}(b)) and Ovito \cite{Stukowski2010} (Fig. \ref{fig:model} and Fig. \ref{fig:mixture_image}(a)). The $g(r)$ functions for determining crystal type were calculated using VMD \cite{Humphrey1996}, and some crystal structure determination was done using pymatgen \cite{Ong2013} and the AFLOW database \cite{Mehl2017a,Hicks2019}.

See Supplemental Material at [URL] for all simulation details. An interactive version of the phase diagrams in Fig. \ref{fig:phase_diagrams} with pair correlation functions can be found at \url{https://aliehlen.github.io/phase_diagrams/}.

\begin{acknowledgments}
The authors would like to thank Martin Girard for useful discussions and review of the work. This work was supported by the Center for Bio-Inspired Energy Science, an Energy Frontier Research Center funded by the US Department of Energy, Office of Science, Basic Energy Sciences under Award DE-SC0000989. H.L.-R. thanks a fellowship from Fulbright-Garcia Robles and A.E. thanks a fellowship from the National Science Foundation under grant DGE-1450006. M.O.d.l.C. thanks the computational support of the Sherman Fairchild Foundation.
\end{acknowledgments}

A.E. and  H.L-R. contributed equally to this work.

\bibliography{refs}

\end{document}